\documentclass[10pt,sort&compress]{article}
\usepackage{graphicx,rotating,multirow}
\usepackage{bm,multirow}
\usepackage{amsmath,amssymb,color,mathrsfs}

\usepackage{footnote}
\usepackage{citesort}

\newcommand{\AJP}{{\em Am. J. Phys.} }

\newcommand{\AO}{{\em Appl. Optics} }

\newcommand{\APB}{{\em Ann. Phys. (Berlin)} }

\newcommand{\CJP}{{\em Can. J. Phys.} }
\newcommand{\CMP}{{\em Commun. Math. Phys.} }

\newcommand{\CPL}{{\em Chem. Phys. Lett.} }
\newcommand{\CQG}{{\em Class. Quantum Grav.} }

\newcommand{\EJP}{{\em Eur. J. Phys.} }

\newcommand{\IJMPA}{{\em Int. J. Mod. Phys. A} }

\newcommand{\JCP}{{\em J. Chem. Phys.} }

\newcommand{\JLTP}{{\em J. Low Temp. Phys.} }

\newcommand{\JMP}{{\em J. Math. Phys.} }
\newcommand{\jpa}{{\em J. Phys. A} }

\newcommand{\NPA}{{\em Nucl. Phys. A} }

\newcommand{\PF}{{\em Phys. Fluids} }

\newcommand{\PLA}{{\em Phys. Lett. A} }
\newcommand{\PR}{{\em Phys. Rev.} }
\newcommand{\PRA}{{\em Phys. Rev. A} }

\newcommand{\PRD}{{\em Phys. Rev. D} }
\newcommand{\PRE}{{\em Phys. Rev. E} }
\newcommand{\PRL}{{\em Phys. Rev. Lett.} }
\newcommand{\PRe}{{\em Phys. Rep.} }

\newcommand{\RMP}{{\em Rev. Mod. Phys.} }
\newcommand{\SPJ}{{\em Sov. Phys. - JETP} }

\newcommand{\ZETF}{{\em Zh. Eksp. Teor. Fiz.} }
\newcommand{\ZP}{{\em Z. Phys.} }
\newcommand{\ZPB}{{\em Z. Phys. B} } 
 
\definecolor{officegreen}{rgb}{0,0.5,0}
\definecolor{pakistangreen}{rgb}{0,0.4,0}
\definecolor{palatinatepurple}{rgb}{0.41,0.16,0.38}
\definecolor{sangria}{rgb}{0.57,0,0.04}

\DeclareMathOperator{\csch}{csch}
\DeclareMathOperator{\sech}{sech} 

\begin{document}
\title{Thermodynamic properties of the one-dimensional Robin quantum well}
\author{O. Olendski\footnote{Department of Applied Physics and Astronomy, University of Sharjah, P.O. Box 27272, Sharjah, United Arab Emirates; E-mail: oolendski@sharjah.ac.ae}}

\maketitle

\begin{abstract}
Thermodynamic properties of the Robin quantum well with extrapolation length $\Lambda$ are analyzed theoretically both for the canonical and two grand canonical ensembles with special attention being paid to the situation when the energies of one or two lowest-lying states are split-off from the rest of the spectrum by the large gap that is controlled by the varying $\Lambda$. For the single split-off level, which exists for the geometry with the equal magnitudes but opposite signs of the Robin distances on the confining interfaces, the heat capacity $c_V$ of the canonical averaging is a nonmonotonic function of the  temperature $T$ with its salient maximum growing to infinity as $\ln^2\Lambda$ for the decreasing to zero extrapolation length and its position being proportional to $1/(\Lambda^2\ln\Lambda)$. The specific heat per particle $c_N$ of the Fermi-Dirac ensemble depends nonmonotonically on the temperature too with its pronounced extremum being foregone on the $T$ axis by the plateau whose value at the dying $\Lambda$ is $(N-1)/(2N)k_B$, with $N$ being a number of the fermions. The maximum of $c_N$, similar to the canonical averaging, unrestrictedly increases as $\Lambda$ goes to zero and is the largest for one particle. The most essential property of the Bose-Einstein ensemble is a formation, for the growing number of bosons, of the sharp asymmetric shape on the $c_N-T$ characteristics that is more protrusive at the smaller Robin distances. This cusp-like structure is a manifestation of the phase transition to the condensate state. For two split-off orbitals, one additional maximum emerges whose position is shifted to the colder temperatures with the increase of the energy gap between these two states and their higher-lying counterparts and whose magnitude approaches $\Lambda$-independent value. All these physical phenomena are qualitatively and quantitatively explained by the variation of the energy spectrum by the Robin distance. Parallels with other structures are drawn and similarities and differences between them are highlighted. Generalization to higher dimensions is provided too.
\end{abstract}

\section{Introduction}\label{sec1}
Thermodynamic properties of the quantum structure are intimately related to its energy spectrum $E_n$, $n=0,1,\ldots$. For example, for canonical ensemble that describes a system that is in thermal equilibrium with much larger bath a basic quantity is the partition function
\begin{equation}\label{Partition1}
Z(\beta)=\sum_ne^{-\beta En},
\end{equation}
where the summation runs over all possible quantum states with the factor $\beta$ being $\beta=1/(k_BT)$, $k_B$ is the Boltzmann constant and $T$ is the thermodynamic temperature of the bath. From this, the mean energy
\begin{subequations}\label{CanonicalMeanEnergy1}
\begin{align}\label{CanonicalMeanEnergy1_1}
\langle E\rangle_{can}(\beta)&=\frac{\sum_{n=0}^\infty E_ne^{-\beta E_n}}{\sum_{n=0}^\infty e^{-\beta E_n}}
\intertext{can equivalently be represented as}
\label{CanonicalMeanEnergy1_2}
\langle E\rangle_{can}&=-\frac{\partial}{\partial\beta}\ln Z.
\end{align}
\end{subequations}
Heat capacity at constant volume $c_V$ is a work that has to be done to change the temperature of the system by one degree
\begin{equation}\label{HeatCapacity1}
c_V(\beta)=\frac{\partial}{\partial T}\langle E\rangle=-k_B\beta^2\frac{\partial}{\partial\beta}\langle E\rangle,
\end{equation}
and for the canonical ensemble it is expressed with the help of the fluctuation-dissipation theorem \cite{Dalarsson1}:
\begin{equation}\label{HeatCapacity2}
c_{can}(\beta)=\beta^2\left(\langle E^2\rangle_{can}-\langle E\rangle_{can}^2\right).
\end{equation}
For $N$ noninteracting particles in the system, the right-hand sides of Eqs.~\eqref{CanonicalMeanEnergy1}--\eqref{HeatCapacity2} have to be multiplied by $N$.

Recent comparative analysis of the one-dimensional (1D) quantum well (QW) with miscellaneous permutations of the Dirichlet and Neumann boundary conditions (BCs) \cite{Olendski22} confirmed that the energy spacing between the orbitals
\begin{equation}\label{Spacing1}
\delta E_n=E_{n+1}-E_n
\end{equation}
plays a crucial role in the thermodynamic properties dependence on temperature; in particular, since this quantity is, at the fixed $n$ (specifically, at $n=0$), the smallest for the pure Neumann structure as compared to other two geometries \cite{Olendski2}, its heat capacity exhibits a salient maximum as a function of $T$ accompanied by the broad minimum at higher temperatures whereas for any other BC configuration the $c_V-T$ characteristics is a smooth line \cite{Olendski22}. Closely related and convenient and useful measure is an energy difference between any excited level and the ground state:
\begin{equation}\label{Spacing2}
\Delta_n=E_n-E_0,\quad n=1,2,\ldots.
\end{equation}
Just this parameter was used for explaining a giant enhancement of the specific heat of the attractive Robin wall in vanishingly small electric fields \cite{Olendski33} when it is practically the same for many quantum orbitals with $n\geq1$ \cite{Olendski3}.

In present research, we apply the methodology developed before \cite{Olendski33,Olendski22} for analyzing thermodynamics of the Robin QW. Robin BC \cite{Gustafson1} for the wave function $\Psi$
\begin{equation}\label{Robin1}
\left.{\bf n}{\bm\nabla}\Psi\right|_{\cal S}=\left.\frac{1}{\Lambda}\Psi\right|_{\cal S},
\end{equation}
with $\bf n$ being an inward unit normal to the surface ${\cal S}$, is characterized by the length $\Lambda$ whose real value guarantees that no current flows through the interface. For the 1D well with the left and right  Robin parameters $\Lambda_-$ and $\Lambda_+$ the energy spectrum is found from the following equation \cite{Olendski1,Olendski44}
\begin{equation}\label{EigenValue1}
\left(\frac{1}{\Lambda_-}+\frac{1}{\Lambda_+}\right)\pi E^{1/2}\cos\pi E^{1/2}+\left(\frac{1}{\Lambda_-\Lambda_+}-\pi^2E\right)\sin\pi E^{1/2}=0.
\end{equation}
Here and below, the distances are measured in units of the well width $d$ and energies -- in units of the ground-state energy of the Dirichlet QW $\pi^2\hbar^2/(2md^2)$, $m$ is a particle mass. This equation shows that for the opposite signs of the equal magnitudes of the extrapolation lengths the energies are \cite{Olendski1,Olendski44}:
\begin{equation}\label{AsymmetricSpectrum1}
E_0=-\frac{1}{\pi^2\Lambda^2},\quad E_n=n^2,\quad\Lambda_-=-\Lambda_+\equiv\Lambda.
\end{equation}
So, it is the Dirichlet spectrum supplemented by the BC split-off state whose  negative energy inversely depends on the square of the Robin distance. The mathematical and physical reasons for this strong binding in general $n$-dimensional domain were explained and analyzed before \cite{Lacey1,Lou1,Asorey1,Levitin1,Berry1,Daners1,Colorado1,Pankrashkin1,Exner1,Freitas1} and repeated for our geometry in a preceding paper \cite{Olendski44} where also quantum-information measures of the structure were computed. Even though it is not a main subject of the present research, let us mention that the geometry from Eq.~\eqref{AsymmetricSpectrum1} does satisfy the  requirement of the Kenneth-Klich theorem \cite{Kenneth1} about the Casimir interaction of the two bodies related by reflection and, accordingly, the Casimir force between the plates will be attractive whereas the theorem does not apply for the surfaces with the same extrapolation lengths considered here too and the corresponding interaction can be repulsive. Below, we show that the increasing gap between negative-energy ground state and its positive counterparts has drastic effects on the thermodynamic properties; namely, it leads to the gigantic enhancement of the heat capacity with its maximum value unrestrectedly increasing with the vanishing $\Lambda$ and simultaneously shifting to the hotter temperatures. For the symmetric QW, $\Lambda_-=\Lambda_+\equiv\Lambda$, at the small negative Robin parameter there are {\em two} split-off orbitals whose energies for the fading extrapolation length are \cite{Olendski44}:
\begin{equation}\label{EnergySymmetricLimit1_MinusZero}
E_{\left\{e,o\right\}}(\Lambda)=-\frac{1}{\pi^2|\Lambda|^2}\left(1\mp4e^{-|\Lambda|^{-1}}\right),\quad\Lambda\rightarrow-0,
\end{equation}
whereas positive-energy levels in the same limit form again the Dirichlet spectrum  from Eq.~\eqref{AsymmetricSpectrum1}. In Eq.~\eqref{EnergySymmetricLimit1_MinusZero} subscripts `$e$` and `$o$` stand for the symmetry of the associated even and odd wave functions, respectively. Emergence of the second split-off state with its energy being exponentially close to the first one brings about new features; namely, interaction between them gives rise to the additional resonance on $c_V-T$ dependence with the finite maximum whose location moves to the zero temperature and whose width shrinks to zero at $|\Lambda|\rightarrow0$. A comparative analysis is performed for the different types of statistical ensembles; in particular, it is shown that for the Fermi-Dirac (FD) distribution the increasing number of particles leads to the decrease of the hot temperature extremum of the specific heat whereas the cold-temperature peak of the symmetric QW exists for one fermion only. For the systems obeying Bose-Einstein (BE) averaging, adding more corpuscles results in the deformation of the resonance shape into the asymmetric cusp-like dependence that is a manifestation of the phase transition and that is accompanied by an almost complete depletion of the ground orbital. Especially promising is the fact that the critical temperature of this transformation from the BE condensate to the normal state can be efficiently controlled by the variation of the Robin extrapolation length and, in particular, can be shifted to the hotter temperatures.

The outline below is as follows. Sec.~\ref{Sec_Asymmetric1} is devoted to the description of the asymmetric QW with SubSec.~\ref{SubSec_AsymmetricCanonical} dealing with the canonical distribution whereas SubSec.~\ref{SubSec_AsymmetricGrandCanonical} introduces initially the features and equations that are common for the two types of the grand canonical ensembles while its SubSubSecs.~\ref{Sec_AsymmetricFermions1} and \ref{SubSec_AsymmetricBE} consider specific properties of fermions and bosons, respectively. The same structure is adopted in Sec.~\ref{Sec_Symmetric1} where a symmetric QW comes under scrutiny with a lot of attention being paid to the cold-temperature resonances for all three types of statistical ensembles. The discussion is wrapped up in Sec.~\ref{Conclusions} by some conclusions.

\section{Asymmetric QW}\label{Sec_Asymmetric1}
Since the energy spectrum of the asymmetric QW is expressed analytically, Eq.~\eqref{AsymmetricSpectrum1}, we start our discussion just from this BC geometry. We will operate with the dimensionless quantities introduced after Eq.~\eqref{EigenValue1}; in addition, below the heat capacity will be measured in units of $k_B$.
\subsection{Canonical ensemble}\label{SubSec_AsymmetricCanonical}
For the asymmetric QW with its spectrum from Eq.~\eqref{AsymmetricSpectrum1}, the partition function and mean energy are:
\begin{eqnarray}\label{CanonicalPartitionAsymmetric1}
Z(\Lambda;\beta)&=&e^{\beta/(\pi^2\Lambda^2)}+\frac{1}{2}\left[-1+\theta_3\left(0,e^{-\beta}\right)\right]\\
\label{CanonicalMeanEnergyAsymmetric1}
\langle E\rangle(\Lambda;\beta)&=&-\frac{\frac{1}{\pi^2\Lambda^2}e^{\beta/(\pi^2\Lambda^2)}+\frac{1}{2}\frac{d\theta_3\left(0,e^{-\beta}\right)}{d\beta}}{e^{\beta/(\pi^2\Lambda^2)}+\frac{1}{2}\left[-1+\theta_3\left(0,e^{-\beta}\right)\right]}.
\end{eqnarray}
Here, $\theta_3(z,q)$ is one of four Theta functions \cite{Bellman1,Abramowitz1}. Note that the expression for heat capacity can be derived analytically from Eqs.~\eqref{HeatCapacity1} and \eqref{CanonicalMeanEnergyAsymmetric1} but it is too bulky and not presented here.
\begin{figure*}
\centering
\includegraphics[width=\columnwidth]{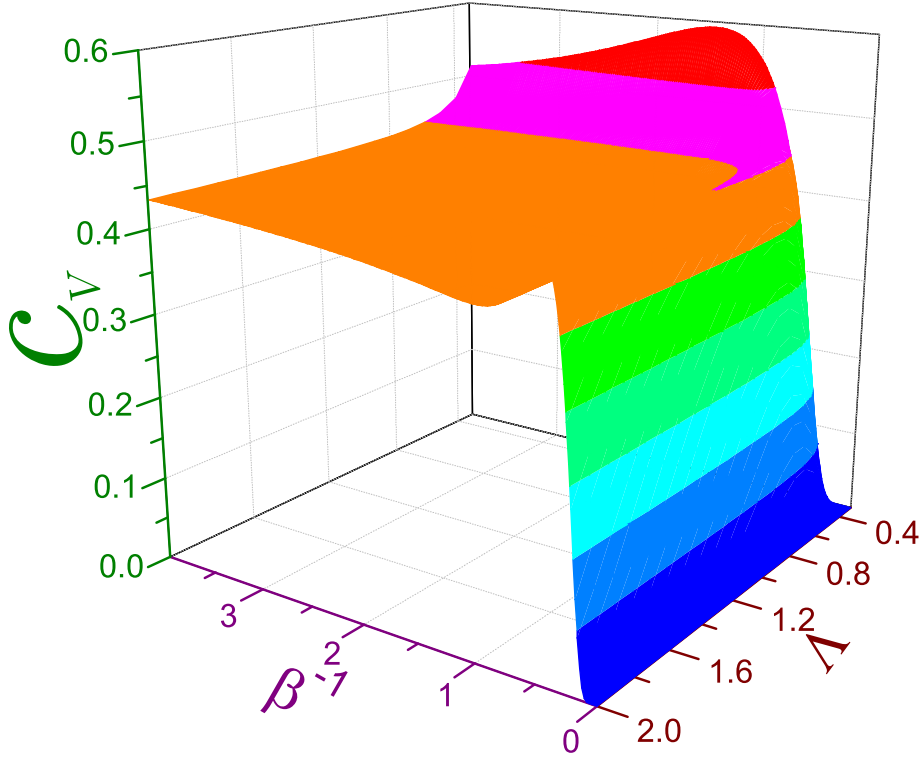}
\caption{\label{AsymmetricCanonicalFig1}Canonical heat capacity $c_V$ of the asymmetric QW as a function of temperature $\beta^{-1}$ and extrapolation length $\Lambda$. Note that the lower limit of the Robin distance is $0.3$. Specific heat dependence on the temperature at the smaller $\Lambda$ is shown in Fig.~\ref{AsymmetricCanonicalFig2}.
}
\end{figure*}

\begin{figure}
\centering
\includegraphics[width=0.78\columnwidth]{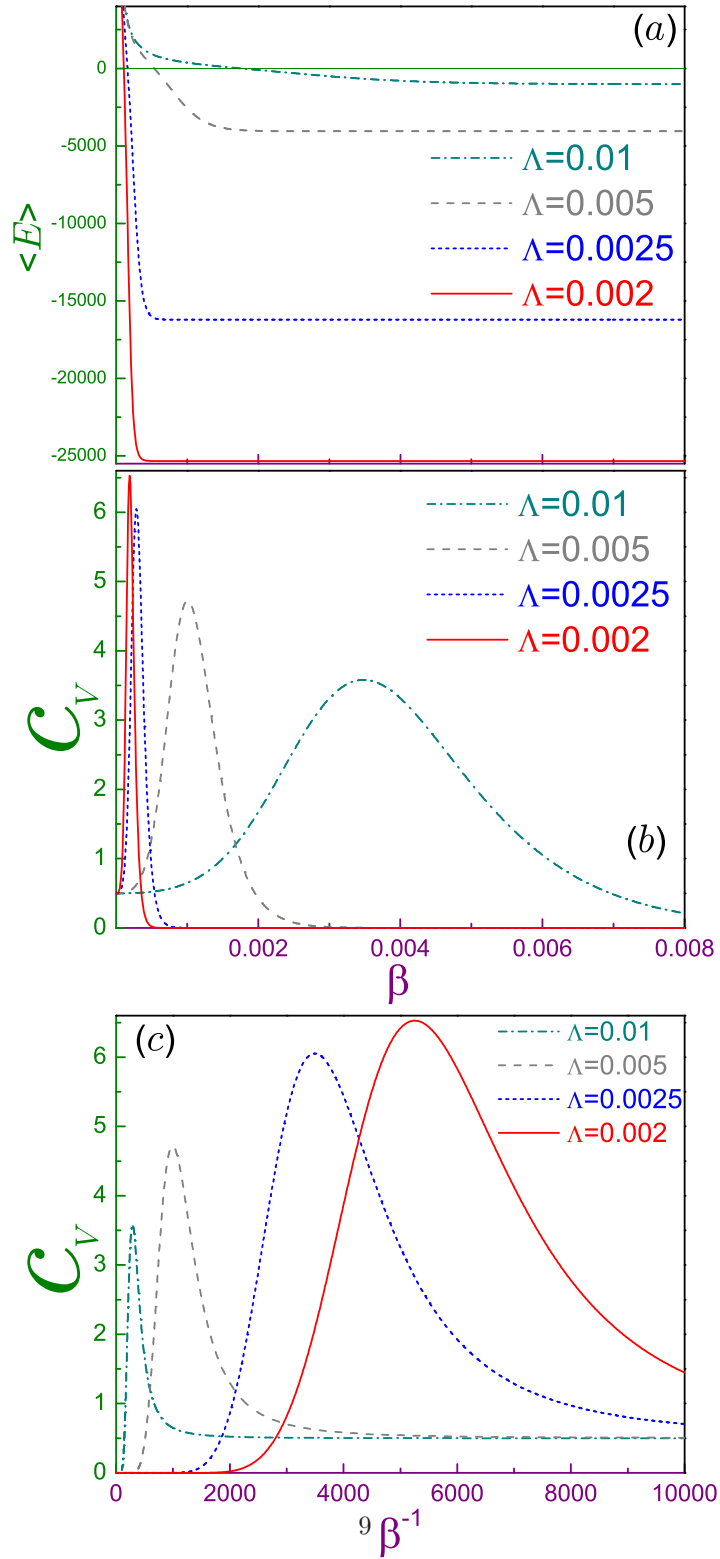}
\caption{\label{AsymmetricCanonicalFig2}Canonical (a) mean energy $\langle E\rangle$ and (b) heat capacity $c_V$ of the asymmetric QW as functions of inverse temperature $\beta$. Panel (c) shows a specific heat dependence on temperature $\beta^{-1}$. In all windows the solid lines correspond to $\Lambda=0.002$, dotted -- to  $\Lambda=0.0025$, dashed curves are for  $\Lambda=0.005$, and dash-dotted ones -- for  $\Lambda=0.01$.
}
\end{figure}

First, consider heat capacity behavior at the cold temperatures , $\beta\rightarrow\infty$, and not very small extrapolation lengths, $\Lambda\gtrsim1$. Then, Eq.~\eqref{CanonicalMeanEnergyAsymmetric1} reduces to
\begin{equation}\label{CanonicalMeanEnergyAsymmetric2}
\langle E\rangle=-\frac{1}{\pi^2\Lambda^2}+\left(1+\frac{1}{\pi^2\Lambda^2}\right)e^{-\gamma}\left(1-e^{-\gamma}+e^{-2\gamma}+\ldots\right),
\end{equation}
and the corresponding specific heat reads:
\begin{equation}\label{CanonicalCapacityAsymmetric1}
c_V=\gamma^2e^{-\gamma}(1-2e^{-\gamma}+3e^{-2\gamma}+\ldots).
\end{equation}
In these equations,
\begin{equation}\label{gamma1}
\gamma\equiv\gamma(\beta,\Lambda)=\beta\left(1+\frac{1}{\pi^2\Lambda^2}\right).
\end{equation}
Note that at $\Lambda=\infty$ Eqs.~\eqref{CanonicalMeanEnergyAsymmetric2} and \eqref{CanonicalCapacityAsymmetric1} transform into their Neumann counterparts \cite{Olendski22}, as expected. It is known that for this limiting BC the heat capacity reaches at $\beta_{max}=2.3031$ ($\beta_{max}^{-1}=0.4342$) a pronounced maximum of $c_{max}=0.4455$ accompanied by the wide minimum at the higher temperatures. Since upon the replacement of $\beta$ by $\gamma$ the structure of Eq.~\eqref{CanonicalCapacityAsymmetric1} stays exactly the same as that for the Neumann QW (see Eq.~(22c) in Ref.~\cite{Olendski22}), one can conclude that the varying Robin distance changes the location of the maximum on the temperature axis as:
\begin{equation}\label{CanonicalBetaMaxAsymmetric1}
\beta_{max}^{-1}(\Lambda)=0.4342\left(1+\frac{1}{\pi^2\Lambda^2}\right)
\end{equation}
without altering the magnitudes of both extrema. Exact numerical results presented in Fig.~\ref{AsymmetricCanonicalFig1} confirm that at large and moderate extrapolation lengths the maximum is indeed shifted to the hotter temperatures as the Robin parameter decreases. This is explained by the widening energy gap between the ground and excited orbitals:
\begin{equation}\label{SpacingAsymmetric1}
\Delta_n(\Lambda)=\frac{1}{\pi^2\Lambda^2}+n^2.
\end{equation}
Magnitude of the maximum stays practically the same for the mentioned above region of $\Lambda\gtrsim1$. However, at the smaller extrapolation distances in addition to quite rapid motion of the extremum to the higher $T$ it also unrestrictedly grows as Robin factor turns to zero. To mathematically describe physical processes that are taking place at the extremely small $\Lambda$, as a first step we use the inversion formula for the Theta function \cite{Olendski22,Bellman1}
\begin{equation}\label{InversionTheta1}
\theta_3\left(0,e^{-\beta}\right)=\left(\frac{\pi}{\beta}\right)^{1/2}\theta_3\!\left(0,e^{-\pi^2/\beta}\right)
\end{equation}
and assuming that the temperature is very high, $\beta\rightarrow0$ (validity of this statement for our range of interest will be justified below), one arrives at the following expression for the mean energy:
\begin{equation}\label{CanonicalMeanEnergyAsymmetric3}
\langle E\rangle=\frac{-\frac{1}{\pi^2\Lambda^2}e^{\beta/(\pi^2\Lambda^2)}+\frac{\pi^{1/2}}{4\beta^{3/2}}}{e^{\beta/(\pi^2\Lambda^2)}+\frac{\pi^{1/2}}{2\beta^{1/2}}},\quad\Lambda\rightarrow0.
\end{equation}
Note that it turns to zero at the temperature $\beta_{\langle E\rangle=0}$, which is found from equation:
\begin{equation}\label{Lambert1}
\frac{1}{\pi^2\Lambda^2}e^{\beta/(\pi^2\Lambda^2)}=\frac{\pi^{1/2}}{4\beta^{3/2}}.
\end{equation}
Its solution is expressed via the Lambert $W$ function\footnote{This function, with its history dating back for more than two and half centuries \cite{Corless1,Hayes1}, in the last thirty years or so has been rapidly coming back from the oblivion due to its miscellaneous applications in physics, astronomy, mathematics and other branches of science \cite{Olendski33,Siewert1,Barry2,Scott3,Scott2,Barry1,Mann1,Mann2,Valluri1,Jenn1,Braun1,Caillol1,Cranmer1,Warburton1,Scott1,Scott5,Shafee1,Scott4,Steinvall1,Valluri2,Pudasaini1,Kamper1,Stewart1,Vial1,Houari1,Houari2,Luo1,Mezo1,Ishkhanyan1,Loudon1,Sacchetti1,Sacchetti2,Roberts1,delaMadrid1,Digilov1}.} \cite{Corless1}:
\begin{equation}
\beta_{\langle E\rangle=0}=\frac{3}{2}\pi^2\Lambda^2W\!\!\left(\frac{16^{2/3}}{24}\frac{1}{\pi^{1/3}\Lambda^{2/3}}\!\right).
\end{equation}
Asymptotics of the Lambert function \cite{Corless1}
\begin{equation}\label{LambertAsymptotics1}
W(x)\rightarrow\ln\frac{x}{\ln x}+\ldots,\quad x\rightarrow\infty,
\end{equation}
shows that at the vanishing Robin parameter the temperature at which the zero mean energy is achieved grows without limits as
\begin{equation}\label{AsymmetricAsymptote1}
\beta_{\langle E\rangle=0}\simeq-\pi^2\Lambda^2\ln\Lambda,\quad\Lambda\rightarrow0.
\end{equation}
This equation manifests that the mean energy for the dying extrapolation lengths becomes a steeper function of the small inverse temperature. This is exemplified in panel (a) of Fig.~\ref{AsymmetricCanonicalFig2} that shows the dependence of the thermally averaged energies on the inverse temperature for several very small $\Lambda$. Rapid growth of $\langle E\rangle$ means, according to Eq.~\eqref{HeatCapacity1}, larger values of the heat capacity whose expression in the same limit reads:
\begin{equation}\label{CanonicalCapacityAsymmetric2}
c_V=\frac{1}{2}\frac{4\beta^{5/2}e^{\beta/(\pi^2\Lambda^2)}+\pi^{9/2}\Lambda^4}{\pi^{7/2}\Lambda^4\left[2\beta^{1/2}e^{\beta/(\pi^2\Lambda^2)}+\pi^{1/2}\right]^2},\quad\Lambda\rightarrow0.
\end{equation}
At infinitely high temperature, $\beta=0$, it degenerates to $1/2$, as expected. Taking a derivative $\partial_\beta$ of Eq.~\eqref{CanonicalCapacityAsymmetric2}, equating it to zero and keeping the largest terms only leads to the following equation for the inverse temperature $\beta_{max}(\Lambda)$ of the maximum of the specific heat:
\begin{equation}\label{AsymmetricAsymptote2}
2\beta^{1/2}e^{\beta/(\pi^2\Lambda^2)}-\pi^{1/2}=0.
\end{equation}
Its solution is:
\begin{equation}\label{AsymmetricAsymptote6}
\beta_{max}(\Lambda)=\frac{1}{2}\pi^2\Lambda^2W\!\!\left(\frac{1}{2\pi\Lambda^2}\right),\quad\Lambda\rightarrow0.
\end{equation}
Invoking the asymptotics from Eq.~\eqref{LambertAsymptotics1} leads essentially to the same dependence (with different small free constant) as for the zero mean energy, Eq.~\eqref{AsymmetricAsymptote1}. Note that in this way our starting assumption of the high temperatures, $\beta\rightarrow0$, is automatically satisfied. Magnitude of the maximum is:
\begin{equation}\label{AsymmetricAsymptote3}
c_{max}(\Lambda)=\frac{1}{4\pi^4\Lambda^4}\beta_{max}^2=\frac{1}{16}W^2\!\left(\frac{1}{2\pi\Lambda^2}\right),\quad\Lambda\rightarrow0,
\end{equation}
or, basically, it is proportional to the square of the logarithm of the extrapolation length:
\begin{equation}\label{AsymmetricAsymptote4}
c_{max}(\Lambda)\sim\ln^2\Lambda,\quad\Lambda\rightarrow0.
\end{equation}
Heat capacity behavior near the extremum is:
\begin{subequations}\label{AsymmetricAsymptote5}
\begin{align}
c_V(\Lambda;\beta)=&c_{max}(\Lambda)-\frac{1}{8\pi^4}\frac{\ln^2\!\Lambda}{\Lambda^4}\left(\beta-\beta_{max}\right)^2,\nonumber\\
\label{AsymmetricAsymptote5_1}
&\left|\beta-\beta_{max}\right|\ll\beta_{max},\quad\Lambda\rightarrow0,
\intertext{or, equivalently:}
c_V\!\left(\Lambda;\beta^{-1}\!\right)=&c_{max}(\Lambda)-\!\left(\!\frac{\pi^4}{8}\Lambda^4\ln^6\!\Lambda\!\!\right)\left(\beta^{-1}-\beta^{-1}_{max}\right)^2,\nonumber\\
\label{AsymmetricAsymptote5_2}
&\left|\beta^{-1}-\beta_{max}^{-1}\right|\ll\beta_{max}^{-1},\quad\Lambda\rightarrow0.
\end{align}
\end{subequations}
Last equation shows that at the smaller extrapolation lengths the resonance gets wider on the $T$ axis. All these features: growth of the maximum value, its shift to the hotter temperatures and widening of the resonance at $\Lambda\rightarrow0$, -- are clearly seen in panels (b) and (c), which depict the heat capacity dependence on the inverse temperature and temperature itself, respectively. It has to be noted also that keeping in Taylor expansion
\begin{eqnarray}
c_V&=&\frac{1}{2}-\frac{2}{\pi^{1/2}}\beta^{1/2}+\frac{6}{\pi}\beta-\frac{2}{\pi^{3/2}}\left(8+\frac{1}{\pi\Lambda^2}\right)\beta^{3/2}+\ldots,\nonumber\\
\label{CanonicalCapacityAsymmetric3}
&&\beta\rightarrow0,\,\Lambda\rightarrow0,
\end{eqnarray}
of Eq.~\eqref{CanonicalCapacityAsymmetric2} first three $\Lambda$-independent terms predicts an existence of the minimum $c_{min}=1/3$ at $\beta_{min}=\pi/36\approx0.08727$ whereas a taking into account of the last item in Eq.~\eqref{CanonicalCapacityAsymmetric3} eliminates this extremum at all. Exact calculations show that the minimum does not exist at $\Lambda\lesssim0.1$.

To explain physically this giant enhancement of the heat capacity described mathematically by Eqs.~\eqref{AsymmetricAsymptote6} -- \eqref{AsymmetricAsymptote5} and exemplified in Fig.~\ref{AsymmetricCanonicalFig2}, one invokes the energy spectrum from Eq.~\eqref{AsymmetricSpectrum1}; namely, as it was stated above, the shrinking extrapolation length leads to the widening gap between split-off negative energy state and its positive counterparts, Eq.~\eqref{SpacingAsymmetric1}. Thus, hotter temperatures are needed to promote the particle from the ground orbital to the first excited level. This explains the shift of the resonance on the $T$ axis to the right at the smaller $\Lambda$. At the same time, the ratio $\Delta_{n+1}(\Lambda)/\Delta_n(\Lambda)$ comes closer to unity:
\begin{equation}\label{AsymmetricAsymptote7}
\frac{\Delta_{n+1}(\Lambda)}{\Delta_n(\Lambda)}=1+(2n+1)\pi^2\Lambda^2\left(1-\pi^2n^2\Lambda^2+\ldots\right),\quad\Lambda\rightarrow0.
\end{equation}
Accordingly, the particle that at $T=0$ resides at the lowest quantum state at the growth of the temperature has about the same probability to make a transition to a huge number of the excited levels. This results in the colossal increase of the heat capacity described mathematically by Eqs.~\eqref{AsymmetricAsymptote3} and \eqref{AsymmetricAsymptote4}. Simultaneously, the width of the resonance increases according to Eq.~\eqref{AsymmetricAsymptote5_2}. It is instructive to compare this phenomenon with the specific heat behavior of the attractive Robin wall in the electric field $\mathscr{E}$ \cite{Olendski33}. In both cases, the resonance unrestrictedly grows with the decrease of either the extrapolation length $\Lambda$, as described above, or the applied voltage \cite{Olendski33} since each of them leads to the asymptotic approach of the ratio $\Delta_{n+1}/\Delta_n$ to unity facilitating in this way the transitions from the ground state to many higher lying levels. However, the diminishing $\mathscr{E}$ leads to the increase of the density of the excited orbitals with the separation between them and their ground counterpart staying practically the same what results in the shift of the maximum to the smaller temperatures and sharpening of the resonance at the vanishing electric fields whereas for the asymmetric Robin QW the decreasing extrapolation length splits stronger the lowest level from higher ones with the difference $E_{n+1}-E_n$ between the positive-energy states, $n=1,2,\ldots$, being unaffected by the $\Lambda$ variation. As a consequence, the location of the peak of the heat capacity moves at the smaller Robin distance to the higher temperatures with the increase of the width of the corresponding resonance.

\subsection{Grand canonical ensemble}\label{SubSec_AsymmetricGrandCanonical}
This type of statistical averaging is applied to open quantum structures that can exchange both heat and matter with the external bath. As a result, the number $N$ of noninteracting corpuscles inside the QW becomes an essential parameter crucially determining its properties. An important thermodynamic quantity of the grand canonical distribution is the chemical potential $\mu$ that is the work that has to be done to add (or remove) one particle to (from) the system
\begin{equation}\label{ChemicalPotential1}
\mu(T,N)=\left(\frac{\partial\langle E\rangle}{\partial N}\right)_{T,V},
\end{equation}
and which is found from
\begin{equation}\label{NumberN_1}
N=\sum_{n=0}^\infty\frac{1}{e^{(E_n-\mu)\beta}\pm1},
\end{equation}
with the plus (minus) sign describing FD (BE) ensemble. Expression for the heat capacity $c_{gc}$ follows from Eq.~\eqref{HeatCapacity1} where the mean energy $\langle E\rangle_{gc}$ is:
\begin{equation}\label{GrandCanonicalMeanEnergy1}
\langle E\rangle_{gc}(\beta,N)=\sum_{n=0}^\infty\frac{E_n}{e^{(E_n-\mu)\beta}\pm1}.
\end{equation}
After some straightforward calculations one gets:
\begin{equation}\label{GrandCanonicalHeatCapacity1}
c_{gc}=\beta^2\sum_{n=0}^\infty\frac{E_n\left(E_n-\mu-\beta\frac{\partial\mu}{\partial\beta}\right)}{\left[e^{(E_n-\mu)\beta}\pm1\right]^2}\,e^{(E_n-\mu)\beta},
\end{equation}
where the derivative of the chemical potential (which for the FD ensemble is also frequently called the Fermi energy) with respect to the inverse temperature reads \cite{Olendski22,Olendski33}:
\begin{equation}\label{Implicit1}
\beta\frac{\partial\mu}{\partial\beta}=\frac{\sum_n\frac{E_n-\mu}{\left[e^{(E_n-\mu)\beta}\pm1\right]^2}\,e^{(E_n-\mu)\beta}}{\sum_n\frac{1}{\left[e^{(E_n-\mu)\beta}\pm1\right]^2}\,e^{(E_n-\mu)\beta}}.
\end{equation}

Asymmetric QW with its spectrum from Eq.~\eqref{AsymmetricSpectrum1} transforms  Eq.~\eqref{NumberN_1} for finding the chemical potential to
\begin{equation}\label{ChemicalPotential2}
N=\frac{1}{e^{-\mu\beta}e^{-\beta\left/\left(\pi^2\Lambda^2\right)\right.}\pm1}+\sum_{n=1}^\infty\frac{1}{e^{-\mu\beta}e^{\beta n^2}\pm1},
\end{equation}
whereas the mean energy becomes:
\begin{equation}\label{GrandCanonicalMeanEnergy2}
\langle E\rangle=-\frac{1}{\pi^2\Lambda^2}\frac{1}{e^{-\mu\beta}e^{-\beta\left/\left(\pi^2\Lambda^2\right)\right.}\pm1}+\sum_{n=1}^\infty\frac{n^2}{e^{-\mu\beta}e^{\beta n^2}\pm1}.
\end{equation}
Similar to the canonical distribution, these equations, which, in general, can be solved only numerically, allow to simplify them in different asymptotic regimes what leads to quite transparent analytic results where now thermodynamic properties, such as heat capacity, in addition to the temperature and Robin parameter, depend on the number of particles too. In line with Subsec.~\ref{SubSec_AsymmetricCanonical}, consider as a first example a situation with not very small extrapolation length, $\Lambda\gtrsim1$, and relatively cold temperatures, $\beta\gtrsim1$. Then, keeping in Eqs.~\eqref{ChemicalPotential2} and \eqref{GrandCanonicalMeanEnergy2} the two lowest levels only, one gets the following equation for finding the chemical potential:
\begin{eqnarray}
&&Ne^{-\mu\beta}e^{-\beta\left/\left(\pi^2\Lambda^2\right)\right.}e^{-2\mu\beta}\nonumber\\
\label{GrandCanonicalAuxEquation1}
&&\pm\left[e^{-\mu\beta}+e^{-\beta\left/\left(\pi^2\Lambda^2\right)\right.}\right](N\mp1)e^{-\mu\beta}+(N\mp2)=0,
\end{eqnarray}
which is correct for one fermion and arbitrary number of bosons. It immediately shows that for one particle the Fermi energy in this regime is locked in between the two states:
\begin{equation}\label{GrandCanonicalAuxEquation2}
\mu_{N=1}^{FD}=\frac{1}{2}\left(1-\frac{1}{\pi^2\Lambda^2}\right),
\end{equation}
and from the expression of the mean energy
\begin{equation}\label{GrandCanonicalAuxEquation3}
\langle E\rangle_{N=1}^{FD}=-\frac{1}{\pi^2\Lambda^2}\frac{1}{e^{-\gamma/2}+1}+\frac{1}{e^{\gamma/2}+1},
\end{equation}
one derives the associated heat capacity:
\begin{equation}\label{GrandCanonicalAuxEquation4}
c_{N=1}^{FD}=\frac{1}{8}\left(\frac{\gamma}{\cosh\frac{\gamma}{4}}\right)^2,\quad\Lambda\gtrsim1,\quad\beta\gtrsim1,
\end{equation}
with $\gamma$ from Eq.~\eqref{gamma1}. Of course, similar to the canonical ensemble, at $\Lambda=\infty$ it degenerates to its Neumann counterpart \cite{Olendski22}. Analysis of this formula that is defered to SubSec.~\ref{Sec_AsymmetricFermions1} reveals that the peak of the heat capacity for the decreasing extrapolation length shifts to the hotter temperatures without changing its value. However, at the very small Robin distances, $\Lambda\rightarrow0$, not only the single-particle extremum is shifted to the larger $T$, but its magnitude also unrestrictedly grows  at $\beta\rightarrow0$. To explain the phenomena taking place in the high-temperature regime at the small $\Lambda$ for both grand canonical ensembles and arbitrary number of corpuscles, one starts from the asymptotic series \cite{Olendski33}:
\begin{subequations}\label{AsymptoticSeries2}
\begin{eqnarray}
&&\sum_{n=0}^\infty\frac{1}{be^{t(n+d)^\alpha}\pm1}=\mp\frac{\Gamma(1+1/\alpha)}{t^{1/\alpha}}{\rm Li}_{1/\alpha}\left(\mp b^{-1}\right)-\frac{d-1/2}{b\pm1}\nonumber\\
\label{AsymptoticSeries2_1}
&&+\frac{b}{1+\alpha}\frac{|d-1/2|^{1+\alpha}}{(b\pm1)^2}t-\ldots,\quad t\rightarrow0\\
&&\sum_{n=0}^\infty\frac{(n+d)^\alpha}{be^{t(n+d)^\alpha}\pm1}=\mp\frac{\Gamma(1+1/\alpha)}{\alpha t^{1+1/\alpha}}{\rm Li}_{1+1/\alpha}\left(\mp b^{-1}\right)\nonumber\\
&&-\frac{1}{1+\alpha}\frac{|d-1/2|^{1+\alpha}}{b\pm1}+\frac{b}{1+2\alpha}\frac{|d-1/2|^{1+2\alpha}}{(b\pm1)^2}t-\ldots,\nonumber\\
\label{AsymptoticSeries2_2}
&&t\rightarrow0,
\end{eqnarray}
\end{subequations}
where $\Gamma(x)$ is $\Gamma$-function \cite{Abramowitz1} and ${\rm Li}_\alpha(x)$ is a polylogarithm \cite{Lewin1}:
\begin{equation}\label{PolyLog1}
{\rm Li}_\alpha(x)=\sum_{k=1}^\infty\frac{x^k}{k^\alpha}.
\end{equation}
Then, at $\beta\rightarrow0$ one has:
\begin{eqnarray}
N&=&\frac{1}{e^{-\mu\beta}e^{-\beta\left/\left(\pi^2\Lambda^2\right)\right.}\!\pm\!1}\!\mp\!\frac{\pi^{1/2}}{2\beta^{1/2}}{\rm Li}_{1/2}\!\left(\mp e^{\beta\mu}\right)\nonumber\\
\label{ChemicalPotential3}
&-&\frac{1}{2}\frac{1}{e^{-\beta\mu}\pm1},\\
\langle E\rangle&=&-\frac{1}{\pi^2\Lambda^2}\frac{1}{e^{-\mu\beta}e^{-\beta\left/\left(\pi^2\Lambda^2\right)\right.}\pm1}\mp\frac{\pi^{1/2}}{4\beta^{3/2}}{\rm Li}_{3/2}\!\left(\mp e^{\beta\mu}\right)\nonumber\\
\label{GrandCanonicalMeanEnergy3}
&-&\frac{1}{24}\frac{1}{e^{-\beta\mu}\pm1}.
\end{eqnarray}
Assuming that the chemical potential is large and negative, we take in these equations the first term only in the Taylor expansion of the polylogarithm:
\begin{eqnarray}\label{ChemicalPotential4}
N&=&\frac{1}{e^{-\mu\beta}e^{-\beta\left/\left(\pi^2\Lambda^2\right)\right.}\!\pm\!1}\!+\!\frac{\pi^{1/2}}{2\beta^{1/2}}e^{\beta\mu}-\frac{1}{2}\frac{1}{e^{-\beta\mu}\pm1},\\
\langle E\rangle&=&-\frac{1}{\pi^2\Lambda^2}\frac{1}{e^{-\mu\beta}e^{-\beta\left/\left(\pi^2\Lambda^2\right)\right.}\!\pm\!1}\!+\!\frac{\pi^{1/2}}{4\beta^{3/2}}e^{\beta\mu}\nonumber\\
\label{GrandCanonicalMeanEnergy4}
&-&\frac{1}{24}\frac{1}{e^{-\beta\mu}\pm1}.
\end{eqnarray}
Eq.~\eqref{ChemicalPotential4} has an analytical solution for $z=e^{\beta\mu}$ (this quantity for the BE ensemble is called the fugacity) but its expression is too unwieldy. Then, to simplify our qualitative analysis, we disregard the last right-hand-side item there to obtain:
\begin{eqnarray}\label{ChemicalPotential5}
e^{-\beta\mu}&=&\frac{1}{4N\beta^{1/2}a}\left[\sqrt{\left[\pi^{1/2}a\mp2(N\mp1)\beta^{1/2}\right]^2\pm8\pi Na\beta^{1/2}}\right.\nonumber\\
&+&\left.\pi^{1/2}a\mp2(N\mp1)\beta^{1/2}\right],
\end{eqnarray}
where, for brevity, a coefficient
$$
a=e^{-\beta\left/\left(\pi^2\Lambda^2\right)\right.}
$$
has been introduced. Taking a Taylor expansion of Eq.~\eqref{ChemicalPotential5} around $\beta=0$ and keeping the first two terms only, one arrives at:
\begin{equation}\label{ChemicalPotential6}
\mu=-\frac{1}{\beta}\ln\!\left(\frac{\pi^{1/2}}{2N\beta^{1/2}}+\frac{1}{N}\right),\quad\beta\rightarrow0.
\end{equation}
Note that the ensemble and extrapolation length dependent factors will appear only starting from the third item of the argument of the logarithm and, since this contribution is proportional to $\beta^{1/2}$, for our analytic description at the hot temperatures can be dropped. Plugging in the leading term of the Fermi energy from Eq.~\eqref{ChemicalPotential6} into Eq.~\eqref{GrandCanonicalMeanEnergy4} (where the last right-hand-side item that is proportional to $\beta^{1/2}$ has been dropped too), the following mean energy is obtained:
\begin{equation}\label{GrandCanonicalMeanEnergy5}
\langle E\rangle=-\frac{1}{\pi^2\Lambda^2}\frac{1}{\frac{\pi^{1/2}}{2N\beta^{1/2}}e^{-\beta\left/\left(\pi^2\Lambda^2\right)\right.}\pm1}+\frac{N}{2\beta},\quad\beta\rightarrow0.
\end{equation}
Then, the heat capacity per particle
\begin{equation}\label{HeatCapN1}
c_N\equiv\frac{c_{gc}}{N}
\end{equation}
in the same limit reads:
\begin{equation}\label{HeatCapN2}
c_N=\frac{1}{2}+\frac{\beta^{3/2}e^{-\beta\left/\left(\pi^2\Lambda^2\right)\right.}\left(2\beta+\pi^2\Lambda^2\right)}{\pi^{7/2}\Lambda^4\left[\pi^{1/2}e^{-\beta\left/\left(\pi^2\Lambda^2\right)\right.}\pm2N\beta^{1/2}\right]^2},\quad\beta\rightarrow0.
\end{equation}
Note that at the infinitely high temperatures, $\beta=0$, it does coincide for both ensembles and is equal to its canonical counterpart. Thus, a well-known independence of the thermodynamic quantities on the type of the statistical distribution at large temperatures \cite{Dalarsson1} is confirmed once again. It is instructive to emphasize that, in order to get analytic results, approximate Eqs.~\eqref{ChemicalPotential6},~\eqref{GrandCanonicalMeanEnergy5} and~\eqref{HeatCapN2} were derived under the assumption of the high temperatures when the chemical potential is large and negative. Upper limit of their applicability is the extremely hot temperature, $\beta^{-1}=\infty$, where they become perfectly accurate. However, since the Robin distance enters into these equations, the lower range where they can be used is strongly $\Lambda$-dependent. By the direct comparison of the exact numerical calculations and those from the above approximate formulae, it is shown in Section~\ref{Sec_AsymmetricFermions1} that in the most important temperature region where  the most interesting phenomena (such as, similar to the canonical ensemble, huge growth of the maximum of the heat capacity) take place, Eqs.~\eqref{ChemicalPotential6},~\eqref{GrandCanonicalMeanEnergy5} and~\eqref{HeatCapN2} become more precise in the limit of the vanishing de Gennes lengths, $\Lambda\rightarrow0$.

\subsubsection{Fermions}\label{Sec_AsymmetricFermions1}
\begin{figure*}
\centering
\includegraphics[width=1.2\columnwidth]{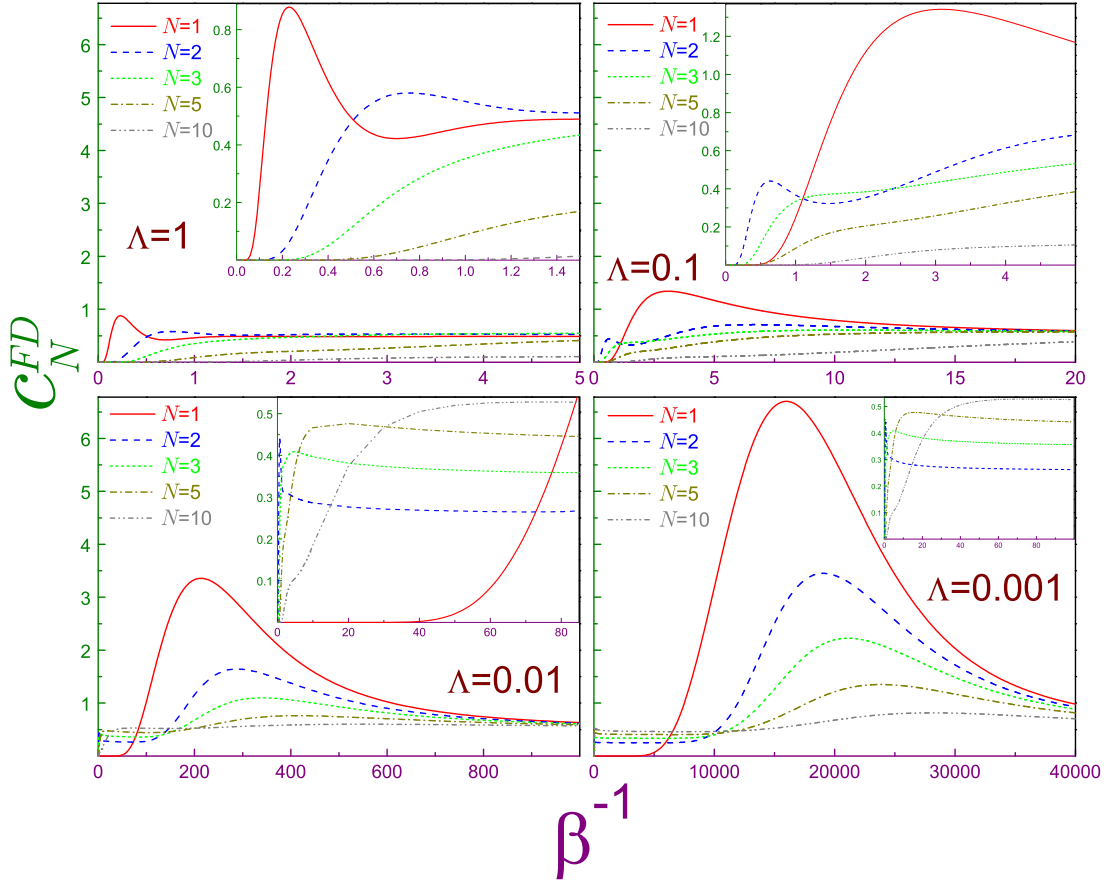}
\caption{\label{AsymmetricFDFig1}
Fermionic heat capacity per particle $c_N^{FD}$ of the asymmetric QW as a function of temperature $\beta^{-1}$ for several extrapolation lengths and number of fermions. Each subplot shows the dependence at the fixed Robin distance $\Lambda$ specified in the corresponding window. Note different temperature ranges for each of the panels. Solid lines are for $N=1$, dashed ones -- for $N=2$, dotted curves -- for $N=3$, dash-dotted ones -- for $N=5$, and the dash-dot-dotted lines depict the heat capacity for $N=10$ fermions. The insets each of which has its own vertical and horizontal scales show enlarged views at the small temperatures where the formation of plateaus from Eq.~\eqref{Plateau1} is demonstrated.
}
\end{figure*}

Quantum particles with half-integer spin do obey the FD statistics. For example, one uses it for the electron with its spin $1/2$. The most characteristic feature of this averaging is the fact that each orbital is occupied by no more than one corpuscle. Armed with Eq.~\eqref{GrandCanonicalAuxEquation4}, we can expand to the finite extrapolation length the previous results for the Neumann QW, $\Lambda=\infty$ \cite{Olendski22}, which predicted the existence of the extremum of the specific heat for one fermion and its absence for any other number of particles, $N\geq2$. These features were explained by the interaction at the very small $T$ of the highest occupied level and its nearest lying above counterpart where a contribution of the lower energy states for two and more fermions makes the heat capacity a quite smooth function of the temperature whereas for the single particle there are no such additional donors that aid to support the continuous growth of the heat capacity that, on the increase of the temperature, reaches maximum and drops. At the quite large Robin lengths, the maximum of the heat capacity
\begin{equation}\label{GrandCanonicalAuxEquation5}
{c_1^{FD}}_{\!\!\!\!\!\!max}=2\csch^2\frac{\gamma_{max}}{4}=0.878,\quad\Lambda\gtrsim1,
\end{equation}
is achieved at $\gamma_{max}=4.799$ ($\gamma_{max}^{-1}=0.208$) that is a solution of equation
\begin{equation}\label{GrandCanonicalAuxEquation6}
\frac{\gamma}{4}\tanh\frac{\gamma}{4}=1.
\end{equation}
In terms of the extrapolation length, the temperature of the maximum reads:
\begin{equation}\label{GrandCanonicalAuxEquation7}
{\beta_1^{FD}}_{\!\!\!\!\!\!max}(\Lambda)=4.799\left(1+\frac{1}{\pi^2\Lambda^2}\right)^{-1},\quad\Lambda\gtrsim1,
\end{equation}
which means that with the decrease of the large and moderate $\Lambda$ the peak is achieved at the higher $T$ what is physically explained by the widening of the energy gap between the ground orbital and the excited levels. At the same time, the half width of the resonance on the temperature axis increases. Simultaneously, for any larger number of corpuscles the smooth $c_N-T$ dependence transforms into one with the more and more conspicuous extremum. For $N=1$ it can be shown from Eq.~\eqref{HeatCapN2} that at the extremely small Robin distances the maximum
\begin{subequations}\label{HeatCap3&4}
\begin{align}\label{HeatCapN3}
{c_1^{FD}}_{\!\!\!\!\!\!max}&=\frac{1}{2}+\frac{1}{4\pi^4\Lambda^4}\left({\beta_1^{FD}}_{\!\!\!\!\!\!max}\right)^2,\quad\Lambda\rightarrow0,
\intertext{is reached at}
\label{HeatCapN4}
{\beta_1^{FD}}_{\!\!\!\!\!\!max}(\Lambda)&=\frac{1}{2}\pi^2\Lambda^2W\!\left(\frac{1}{2\pi\Lambda^2}\right),\quad\Lambda\rightarrow0.
\end{align}
\end{subequations}
Note that this temperature is exactly the same as its canonical counterpart, Eq.~\eqref{AsymmetricAsymptote6}, and the corresponding peak values, Eqs.~\eqref{HeatCapN3} and \eqref{AsymmetricAsymptote3}, are also practically equal. This is not surprising since, as was mentioned above, at the high temperatures the difference between the canonical and grand canonical ensembles is negligible \cite{Dalarsson1}.

Fig.~\ref{AsymmetricFDFig1} shows evolution of the heat capacity-temperature dependence with the varying extrapolation length. Tracking of the solid curve in different panel shows a transformation from the near-Neumann regime, Eqs.~\eqref{GrandCanonicalAuxEquation5} and \eqref{GrandCanonicalAuxEquation7}, to the ultra Robin one, Eqs.~\eqref{HeatCap3&4}. The maximum of the specific heat for the single fermion, $N=1$, gets larger at the smaller $\Lambda$ and is shifted to the hotter $T$, as Eqs.~\eqref{HeatCap3&4} manifest. The resonance is also widening, as can be shown by calculating a second derivative of Eq.~\eqref{HeatCapN2} at ${\beta_1^{FD}}_{\!\!\!\!\!\!max}(\Lambda)$ from Eq.~\eqref{HeatCapN4}. A physical explanation of these phenomena is the same as for the canonical ensemble discussed in the previous section.

To show the convergence of the approximate Eqs.~\eqref{ChemicalPotential6}, \eqref{GrandCanonicalMeanEnergy5} and~\eqref{HeatCapN2} to the exact numerical results, we provide the temperatures at which the maximum of the specific heat is achieved; for example, at $\Lambda=0.001$, Eq.~\eqref{HeatCapN4} yields $\beta_{max}^{-1}=2.09\times10^4$ ($\beta_{max}=4.79\times10^{-5}$) whereas the exact one shown in Fig.~\ref{AsymmetricFDFig1} is equal to $\beta_{max}^{-1}=1.60\times10^4$ ($\beta_{max}=6.25\times10^{-5}$) with their ratio being $1.31$ ($0.763$). The same quantities for $\Lambda=0.0001$ are, respectively: $\beta_{max}^{-1}=1.45\times10^6$ ($\beta_{max}=6.88\times10^{-7}$), $\beta_{max}^{-1}=1.24\times10^6$ ($\beta_{max}=8.05\times10^{-7}$), and $1.17$ ($0.855$). These data manifest that at the smaller Robin distances the relative difference between the exact numerical and approximate analytic calculations does decrease.

Similar to the attractive Robin wall in the vanishing electric field \cite{Olendski33}, the increasing number of the electrons in the system subdues the peak and moves it to the warmer temperatures since the higher lying at $T=0$ fermions impede the interaction of their lowest counterpart with the levels from the positive part of the spectrum. Mathematically, this immediately follows from Eq.~\eqref{HeatCapN2} where the number of particles enters quadratically into the denominator. At the number of corpuscles greater than one, there are two other remarkable features worth mentioning. First, as the insets in the lower panels of Fig.~\ref{AsymmetricFDFig1} demonstrate, at the small and ultra small extrapolation lengths, the structure with the two fermions exhibits an additional maximum whose peak value at $\Lambda\rightarrow0$ tends to $0.441$ with its location approaching $\beta_{max}^{-1}=0.633$. It is easy to explain the formation of this finite strength resonance by taking into account that at such Robin distances the gap $\Delta_1(\Lambda)$ is huge, see Eq.~\eqref{SpacingAsymmetric1}, and, accordingly, the particle occupying the lowest energy state does not contribute at the small and moderate temperatures to the heat capacity. Then, in this regime one can safely consider the structure as the Dirichlet QW with $N-1$ electrons for which, as was shown earlier \cite{Olendski22}, single fermion creates a heat capacity resonance with the peak of $0.882$ reached at just above mentioned temperature whereas for any larger number $N$ the $c_N-T$ characteristics is a monotonic function. Recalling that the actual number of electrons in the Robin QW is one unit greater, the dependence shown in the insets is obtained. Second, the high-temperature resonance is preceded on the $T$ axis by the temperature-independent plateau with its value at the vanishing $\Lambda$ equal to
\begin{equation}\label{Plateau1}
c_{N_{pl}}=\frac{1}{2}\frac{N-1}{N}.
\end{equation}
Explanation of these flat regions is similar to the previous geometry \cite{Olendski33}; namely, at $T=0$ one particle occupies the orbital with the negative energy from Eq.~\eqref{AsymmetricSpectrum1} whereas remaining $N-1$ fermions consecutively fill the levels of the Dirichlet spectrum. With the growth of the temperature, each of these positive-energy electrons attains the classical heat capacity of $1/2$ at the relatively moderate $T$ but the large gap $\Delta_1(\Lambda)$ forbids in this temperature regime any contribution to the total specific heat from the negative-energy corpuscle. Only when the strength of the thermal quantum becomes comparable to the energy difference $\Delta_1(\Lambda)$, it starts to contribute to the specific heat producing the $N$ dependent resonance. Since the gaps $\Delta_n(\Lambda)$ for the considered geometry increase at the shrinking extrapolation length, the widths of the plateaus on the $T$ axis get larger for the smaller vanishing $\Lambda$.

\subsubsection{Bosons}\label{SubSec_AsymmetricBE}
For the particles described by the BE statistics, any arbitrary number of corpuscles can coexist in the same quantum state; in particular, they, under special conditions, form a condensate when the overwhelming majority of bosons reside on the ground level. Theoretical prediction \cite{Bose1} of this remarkable property of the bosonic collective motion and its impressive first experimental confirmation \cite{Anderson1,Bradley1,Davis1} are separated in time by seventy years. The latter rekindled the huge interest in the study of BE systems \cite{Dalfovo1,Pitaevskii1,Leggett1,Pethick1}. As a prerequisite to the analysis below, we would like to point out on the theoretical discussion of the influence of the miscellaneous forms of the confining potential \cite{Olendski33,Dalfovo1,Ketterle1,Druten1,Napolitano1,Mullin1,Bagnato1,Bagnato2} and/or BCs \cite{Olendski22,Olendski33,Mills1,Krueger1,Sonin1,Greenspoon1,Zasada1,Goble1,Goble2,Goble3,Pathria1,Greenspoon2,Barber1,Hasan1,Robinson1,Ziff1,Landau1,VanDenBerg1,Grossmann1} (including Robin ones \cite{Robinson1,Landau1,VanDenBerg1}) on the properties of the noninteracting bosons.
\begin{figure*}
\centering
\includegraphics[width=1.2\columnwidth]{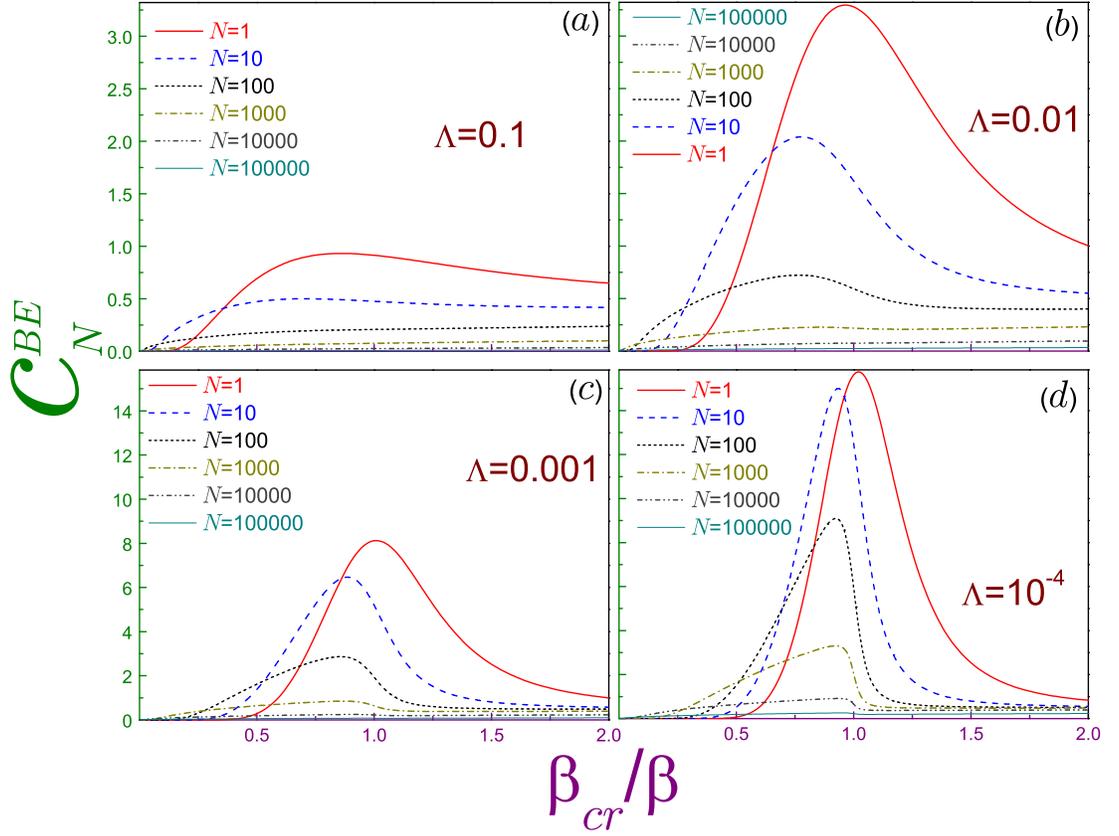}
\caption{\label{AsymmetricBEFig1}Bosonic heat capacity per particle $c_N$ of the asymmetric well as a function of the normalized temperature $\beta_{cr}/\beta$ for (a) $\Lambda=0.1$, (b) $\Lambda=0.01$, (c) $\Lambda=0.001$ and (d) $\Lambda=10^{-4}$ and several numbers $N$ of bosons where thick solid lines are for $N=1$, dashed curves -- for $N=10$, dotted lines -- for $N=100$, dash-dotted curves -- for $N=1000$, dash-dot-dotted ones are for $N=10000$, and thin solid lines -- for $N=100000$. Note different vertical ranges for the lower and upper panels.
}
\end{figure*}

In the physics of the boson thermodynamics, in addition to heat capacity, two other important physical quantities are used: ground-state population $n_0$ and critical temperature $\beta_{cr}^{-1}$. The former represents a ratio of the number of particles $N_0$ on the lowest-energy level
\begin{equation}\label{Number_N0}
N_0=\frac{1}{e^{(E_0-\mu)\beta}-1}
\end{equation}
to the total number of bosons in the well:
\begin{equation}\label{Number_n0}
n_0(\Lambda;\beta)=\frac{N_0}{N}.
\end{equation}
Since this quantitative means of the BE condensation depends on $\beta^{-1}$, it is relevant to introduce the critical temperature $\beta_{cr}^{-1}$ as the largest temperature at which the BE condensation still persists; namely, it is a situation with zero particles occupying the lowest level, $N_0=0$, and with the chemical potential being locked onto the ground-state energy, $\mu=E_0$ \cite{Ketterle1}:
\begin{equation}\label{CriticalTemp1}
\sum_{n=1}^\infty\frac{1}{e^{(E_n-E_0)\beta_{cr}}-1}=N.
\end{equation}
Critical temperature grows for the larger number of bosons and decreases with Robin distance increasing. Under the assumption of the small extrapolation length, which was also used in derivation of Eqs.~\eqref{ChemicalPotential3}--\eqref{HeatCapN2}, one finds for the asymmetric QW:
\begin{subequations}\label{CriticalTemp2}
\begin{align}\label{CriticalTemp2_1}
\beta_{cr}=\frac{\pi^2}{2}\Lambda^2W\!\!\left(\frac{1}{2\pi^2\Lambda^2N^2}\right),\quad\Lambda\rightarrow0,
\intertext{what, basically, becomes:}
\label{CriticalTemp2_2}
\beta_{cr}=-\frac{\pi^2}{2}\Lambda^2\ln\!\left(2\pi^2\Lambda^2N^2\right),\quad\Lambda\rightarrow0.
\end{align}
\end{subequations}
\begin{figure*}
\centering
\includegraphics[width=1.2\columnwidth]{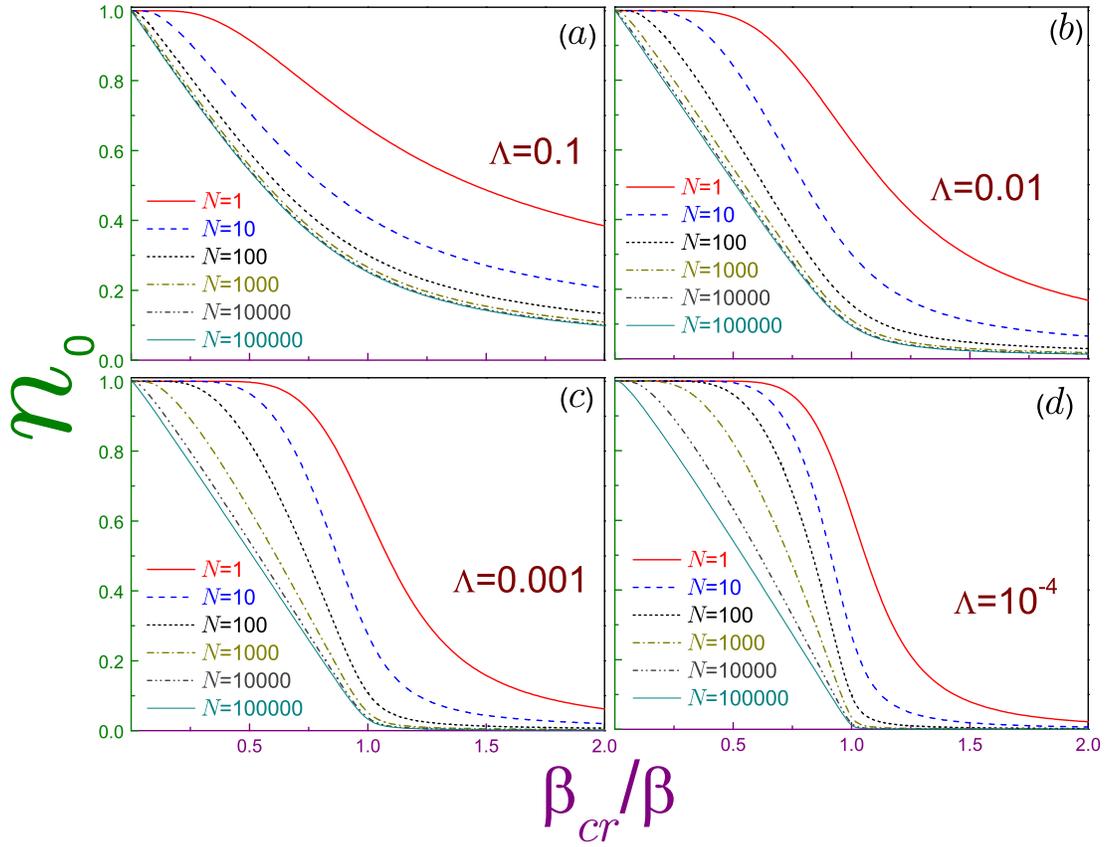}
\caption{\label{AsymmetricBEFig2}Ground-state occupation $n_0$ of the asymmetric QW as a function of the temperature normalized in units of $T_{cr}$. The same conventions as in Fig.~\ref{AsymmetricBEFig1} are used.
}
\end{figure*}
\begin{figure}
\centering
\includegraphics[width=\columnwidth]{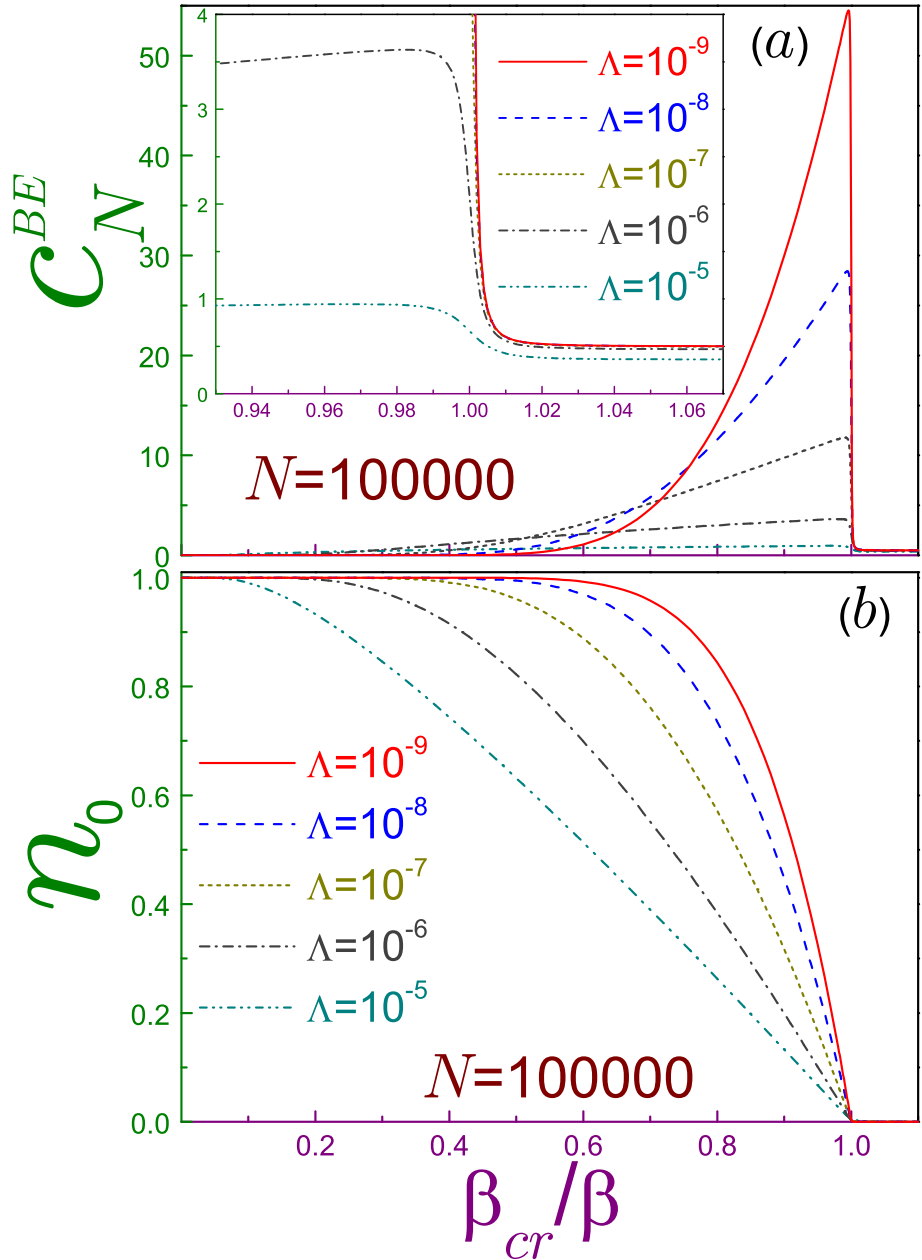}
\caption{\label{AsymmetricBEFig3}Bosonic (a) heat capacity per particle $c_N^{BE}$ and (b) ground-state occupation $n_0$ of the asymmetric QW as a function of the normalized temperature for several extrapolation lengths and fixed number of particles $N=10^5$. Solid lines denote dependencies for $\Lambda=10^{-9}$, dashed curves -- for $\Lambda=10^{-8}$, dotted ones -- for $\Lambda=10^{-7}$, dash-dotted lines are for $\Lambda=10^{-6}$, and dash-dot-dotted curves -- for $\Lambda=10^{-5}$. Inset in panel (a) shows an enlarged view near the critical temperature.
}
\end{figure}

Fig.~\ref{AsymmetricBEFig1} exhibits BE heat capacity per particle in terms of $T/T_{cr}$ for several Robin parameters and numbers of bosons. It was shown before that for any permutation of the Neumann and Dirichlet BCs the heat capacity is a monotonic function of the temperature at the arbitrary $N$ \cite{Olendski22}. Our analysis reveals that the same holds true for the extrapolation lengths $\Lambda\gtrsim1$. At the smaller distances the maximum of the heat capacity starts to form with this process taking place first for the lesser number of bosons; for example, panel (a) exhibits shallow and broad but conspicuous extrema for one and ten particles whereas for the larger $N$ the $c_N(T)$ dependence remains a monotonic function changing from zero at $T=0$ to $1/2$ for the very hot environment, $T\rightarrow\infty$. The decrease of the extrapolation length increases the gap $\Delta_n(\Lambda)$ and, accordingly, causes the growth of the maximum of $c_N$, as a comparison of the panels of Fig.~\ref{AsymmetricBEFig1} shows. Not shown in the figure is the fact that, similar to the attractive Robin wall in the electric field \cite{Olendski22}, for moderate number of particles, $N\lesssim10$, the peak value might be a $\Lambda$-dependent increasing function of $N$ whereas for the larger systems the maximum decreases with the growing $N$. Simultaneously, as the crowd of bosons gets bigger, the BE statistics begins to dominate by transforming this geometrically induced resonance, which, at least, near its maximum, is a symmetric function of the difference $\beta^{-1}-\beta_{max}^{-1}$, into the highly asymmetric cusp-like shape with its right part becoming more upright for the increasing $N$ and decreasing $\Lambda$. Some hints on the formation of this structure can already be seen in panel (b) of Fig.~\ref{AsymmetricBEFig1} and they become more and more salient as the extrapolation length decreases, panels (c) and (d). Note that in the same limit of vanishing Robin parameter, $\Lambda\rightarrow0$, the location of the maximum on $T$ axis and subsequent rapid descent of the capacity come closer and closer to the critical temperature $T_{cr}$. At the same time, ground-state population in the interval $0\leq T\leq T_{cr}$ becomes a steeper function of the temperature  almost turning to zero at $\beta_{cr}^{-1}$, as a comparison of different panels of Fig.~\ref{AsymmetricBEFig2} demonstrates. At zero temperature, no boson is found on any of the excited levels, $\left. n_0\right|_{T=0}=1$. The warming of the QW expels them from the lowest orbital. To understand the ground-state population dependence on temperature at extremely small $T$ and at the large gap between the ground and first excited levels, one first finds the corresponding chemical potential:
\begin{equation}\label{ChemicalPotential7}
\mu=-\frac{1}{\pi^2\Lambda^2}-\frac{1}{\beta}\ln\!\left(1+\frac{1}{N-e^{-\beta/(\pi^2\Lambda^2)}}\right),\quad\Lambda\rightarrow0,\quad\beta\rightarrow\infty.
\end{equation}
Then, the relative number of particles in the lowest state is obtained as:
\begin{subequations}\label{Number_n0_2}
\begin{align}\label{Number_n0_2_1}
n_0&=1-\frac{1}{N}\,e^{-\beta/(\pi^2\Lambda^2)},\quad\Lambda\rightarrow0,\quad\beta\rightarrow\infty,
\intertext{or, recalling the approximate expression for the critical temperature, Eq.\eqref{CriticalTemp2_2}:}
\label{Number_n0_2_2}
n_0&=1-\left(2^{1/2}\pi\Lambda N\right)^{\beta/\beta_{cr}},\quad\Lambda\rightarrow0,\quad\beta\rightarrow\infty.
\end{align}
\end{subequations}
Near the phase transition, a substitution of the chemical potential from Eq.~\eqref{ChemicalPotential6} into Eq.~\eqref{NumberN_1} yields:
\begin{equation}\label{Number_n0_3}
\left. n_0\right|_{\beta=\beta_{cr}}=\frac{1}{N}\left[-\frac{2}{\pi}\ln\!\left(2^{1/2}\pi\Lambda N\right)\right]^{1/2},\quad\Lambda\rightarrow0.
\end{equation}
Eqs.~\eqref{Number_n0_2_2} and \eqref{Number_n0_3} confirm qualitatively the dependencies shown in Figs.~\ref{AsymmetricBEFig2} and \ref{AsymmetricBEFig3}(b), which express the fact that the decrease of $n_0$ at  $T\ll T_{cr}$ ($T\sim T_{cr}$) is flatter (more  precipitous) at the smaller Robin lengths assuming that the temperature is measured in units of its critical counterpart. Eq.~\eqref{Number_n0_2_1} also manifests that at the same cold temperature the ground-state population deviates less from unity for the bigger flock of corpuscles whereas in terms of $T/T_{cr}$ the corresponding curve becomes, for the larger $N$, a steeper function of this small ratio, as exemplified in Fig.~\ref{AsymmetricBEFig2}. Also, as it follows from Eq.~\eqref{Number_n0_3} and is shown in Fig.~\ref{AsymmetricBEFig2}, at the critical temperature the ground-state population gets smaller for the larger $N$. For the infinite number of bosons, $N=\infty$, the lowest orbital will be completely depopulated at $T\geq T_{cr}$ what means a full destruction of the BE condensate by the heating of the structure but due to the finiteness of $N$ a tiny $\Lambda$-dependent fraction $n_0$ persists for the temperatures above the critical one. To underline the significance of the varying extrapolation lengths for the evolution of BE condensate, Fig.~\ref{AsymmetricBEFig3} shows specific heat $c_N^{BE}$ and ground-state occupation $n_0$ at the fixed number of bosons $N=100000$ and several very small Robin distances. The decreasing $\Lambda$ leads to a  growth of the maximum of the specific heat (which was also the case for the canonical and FD distributions) while the depopulation of the ground orbital occurs at the larger $T/T_{cr}$. For the dying extrapolation length, the ground-state occupation asymptotically transforms into the step function $h(x)=\left\{\begin{array}{cc}
1,&x\geq0\\
0,&x<0
\end{array}\right.$:
\begin{subequations}\label{AsymptoteBose1}
\begin{align}\label{AsymptoteBose1_n0}
n_0(T)&\xrightarrow[\Lambda\rightarrow0]{}h(T_{cr}-T),
\intertext{whereas the critical temperature itself tends to infinity [see Eqs.~\eqref{CriticalTemp2}]:}
\label{AsymptoteBose1_Tcr}
T_{cr}&\xrightarrow[\Lambda\rightarrow0]{}\infty.
\end{align}
\end{subequations}
This makes the difference with the previously considered attractive Robin wall \cite{Olendski33} where the increase of the maximum of the heat capacity by the variation of the electric field is accompanied by its simultaneous shift to the colder temperatures. Thus, experimental realization of the bosonic Robin QW with the attractive surface will allow to increase the critical temperature at will. Note that Eq.~\eqref{CriticalTemp2_1} for $T_{cr}$ was obtained straightforwardly from Eq.~\eqref{CriticalTemp1}. Another method to arrive at it is to zero the denominator in Eq.~\eqref{HeatCapN2} what results in infinite specific heat at the transition point. Taking into account higher-order terms that were neglected in the derivation of the latter formula leads to the finite $\Lambda$-dependent maximum, as exact results from Fig.~\ref{AsymmetricBEFig3}(a) show. At the infinite number of bosons, $N=\infty$, the cusp-like shape at $T=T_{cr}$ becomes a discontinuity disclosing in this way a phase transition; namely, in this particular situation, it is a transition from the BE condensate to the normal phase of the noninteracting corpuscles in the asymmetric Robin QW. This highly asymmetric cusp-like shape of the specific heat was intensively analyzed theoretically \cite{Olendski3,Olendski33,Pitaevskii1,Pethick1,Druten1,Napolitano1,Grossmann2,Haugerud1,Haugset1,Goswami1} and demonstrated experimentally \cite{Ensher1}. For the single Robin wall with the negative extrapolation length, it was shown that the sharpness of this feature can be effectively controlled by the applied voltage \cite{Olendski33}. The results above confirm that the same holds true for the Robin parameter of the asymmetric QW.

\section{Symmetric QW}\label{Sec_Symmetric1}
For the geometry with the same BC at each surface, the cases of the positive and negative extrapolation lengths should be considered separately. For the Robin distance $\Lambda$ changing from zero to large positive values, one observes a continuous transformation from the Dirichlet features to the Neumann ones. Since the split-off states are created for the negative $\Lambda$ only, we concentrate below just on this configuration. For the symmetric QW, an existence at small negative extrapolation distances of the {\em two} split-off levels with almost equal negative energies, Eq.~\eqref{EnergySymmetricLimit1_MinusZero}, produces, in addition to the extrema discussed in Sec.~\ref{Sec_Asymmetric1}, an extra resonance. As is shown below, for either type of statistics it at $\Lambda\rightarrow-0$ shifts to colder temperatures with its maximum value being almost intact by the varying Robin distance and its half-width rapidly decreasing.
\subsection{Canonical ensemble}\label{SubSec_SymmetricCanonical}
\begin{figure}
\centering
\includegraphics[width=\columnwidth]{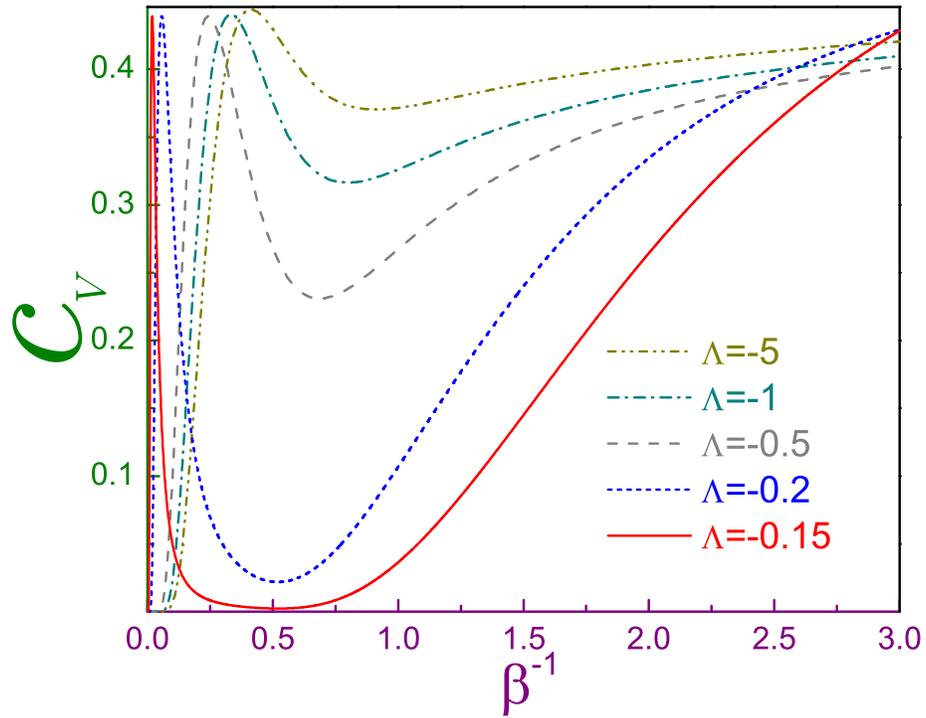}
\caption{\label{SymmetricCanonicalFig1}Canonical heat capacity $c_V$ of the symmetric QW as function of normalized temperature $\beta^{-1}$. Solid line corresponds to $\Lambda=-0.15$, dotted -- to  $\Lambda=-0.2$, dashed curve is for  $\Lambda=-0.5$, dash-dotted one -- for  $\Lambda=-1$, and dash-dot-dotted lines is for $\Lambda=-5$.
}
\end{figure}

As Fig.~\ref{SymmetricCanonicalFig1} demonstrates, the resonance that existed at $\Lambda=-\infty$ \cite{Olendski22} moves to the left without basically changing its magnitude as $|\Lambda|$ decreases. Simultaneously, the accompanying minimum deepens and at quite small extrapolation parameter reaches zero. To understand this behavior, it suffices to retain in the expression for the mean energy, Eq.~\eqref{CanonicalMeanEnergy1_1}, only the terms from Eq.~\eqref{EnergySymmetricLimit1_MinusZero}:
\begin{equation}\label{CanonicalMeanEnergySymmetric1}
\langle E\rangle=-\frac{1}{\pi^2\Lambda^2}\left[1+4e^{-|\Lambda|^{-1}}\tanh\!\!\left(\frac{4\beta}{\pi^2|\Lambda|^2}e^{-|\Lambda|^{-1}}\right)\right].
\end{equation}
Corresponding heat capacity reads:
\begin{equation}\label{CanonicalCapacitySymmetric1}
c_V=16\frac{\beta^2}{\pi^4|\Lambda|^4}e^{-2|\Lambda|^{-1}}\!\cosh\!\left(\frac{4\beta}{\pi^2|\Lambda|^2}e^{-|\Lambda|^{-1}}\right)^{-2},
\end{equation}
which is valid at $\Lambda\rightarrow-0,\,\beta\rightarrow\infty$. By taking a derivative of this expression with respect to $\beta$, one finds that the $\Lambda$-independent maximum
\begin{equation}\label{SymmetricCanonicalAsymptote2}
c_{max}=\left(\frac{s}{\sinh s}\right)^2=0.4392
\end{equation}
is located at
\begin{equation}\label{SymmetricCanonicalAsymptote1}
\beta_{max}=\frac{\pi^2}{4}|\Lambda|^2e^{|\Lambda|^{-1}}s,\quad\Lambda\rightarrow-0,
\end{equation}
where $s=1.1997$ is a solution of equation
\begin{equation}\label{SymmetricEquation1}
x=\coth x.
\end{equation}
The behavior near this extremum reads:
\begin{subequations}\label{SymmetricCanonicalAsymptote3}
\begin{align}
c_V(\Lambda;\beta)=&c_{max}-s_1|\Lambda|^{-4}e^{-2|\Lambda|^{-1}}\left(\beta-\beta_{max}\right)^2,\nonumber\\
\label{SymmetricCanonicalAsymptote3_1}
&\left|\beta-\beta_{max}\right|\ll\beta_{max},\,\Lambda\rightarrow-0,
\intertext{or, equivalently:}
c_V\!\left(\Lambda;\beta^{-1}\!\right)=&c_{max}-\frac{\pi^8}{256}s^4s_1|\Lambda|^4e^{2|\Lambda|^{-1}}\left(\beta^{-1}-\beta_{max}^{-1}\right)^2,\nonumber\\
\label{SymmetricCanonicalAsymptote3_2}
&\left|\beta^{-1}-\beta_{max}^{-1}\right|\ll\beta_{max}^{-1},\,\Lambda\rightarrow-0.
\end{align}
\end{subequations}
Here,
\begin{eqnarray}
s_1&=&\frac{8}{\pi^4}\frac{4s\sinh2s+4s^2-2s^2\cosh2s-1-\cosh2s}{\cosh^4s}\nonumber\\
\label{s_1}
&=&0.07215.
\end{eqnarray}
Eq.~\eqref{SymmetricCanonicalAsymptote3_2} shows that the resonance rapidly sharpens on the $T$ axis for the decreasing $|\Lambda|$ with its simultaneous shift to colder temperatures, as it follows from Eq.~\eqref{SymmetricCanonicalAsymptote1}. Both these features are vividly seen in Fig.~\ref{SymmetricCanonicalFig1}. Extrapolation length can not alter the maximal value of the resonance from Eq.~\eqref{SymmetricCanonicalAsymptote2} since in the regime $\Lambda\rightarrow-0$, $T\rightarrow0$ it is formed by the interaction of the two split-off levels only without a contribution from the higher-lying orbitals but since the energy difference between these two states
\begin{equation}\label{SymmetricCanonicalAsymptote4}
\Delta_1(\Lambda)=\frac{8}{\pi^2|\Lambda|^2}e^{-|\Lambda|^{-1}},\quad\Lambda\rightarrow-0,
\end{equation}
rapidly shrinks, the smaller temperatures are needed to achieve this extremum what results in its shift towards $T=0$.

\begin{figure}
\centering
\includegraphics[width=\columnwidth]{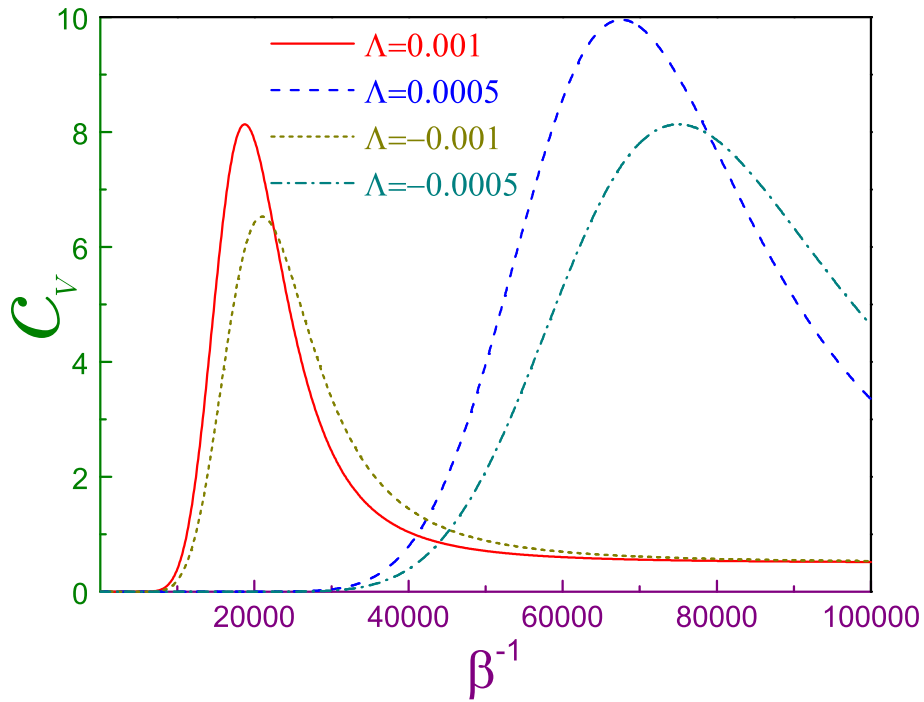}
\caption{\label{SymmetricCanonicalFig2}Comparison of canonical heat capacities $c_V$ of the symmetric and asymmetric QWs at two small absolute values of the extrapolation length. Positive values of the Robin distances ($\Lambda=0.001$ for the solid line and $\Lambda=0.0005$ for the dashed curve) correspond to the asymmetric structure whereas the negative ones ($\Lambda=-0.001$ for the dashed and $\Lambda=-0.0005$ for the dash-dotted lines) are those describing a symmetric geometry. Due to their extremely narrow widths, resonances at the cold temperatures described by Eqs.~\eqref{SymmetricCanonicalAsymptote2} -- \eqref{s_1} are not resolved for the horizontal scale of the figure.
}
\end{figure}

Similar to the asymmetric QW, the structure with equal extrapolation lengths possesses at the small negative Robin distances a pronounced high-temperature maximum of the specific heat with its magnitude increasing at the vanishing $\Lambda<0$. Its physical explanation is identical to the one provided in SubSec.~\ref{SubSec_AsymmetricCanonical} with the same qualitative but slightly different quantitative features; for example, the mean energy at $\Lambda\rightarrow-0$ is described by the modified form of Eq.~\eqref{CanonicalMeanEnergyAsymmetric3} where the first items in the numerator and denominator of the right-hand side have to be multiplied by the factor of two. This leads, as Fig.~\ref{SymmetricCanonicalFig2} demonstrates, to a decrease of the peak of the heat capacity and its shift to the warmer temperatures.

\subsection{Grand canonical ensemble}
Under the same assumption as the one used in the previous subsection, in Eq.~\eqref{NumberN_1} again only the two terms associated with the lowest energies from Eq.~\eqref{EnergySymmetricLimit1_MinusZero} can be considered what results in
\begin{equation}\label{NumberN_2}
Nx^2\pm2(N\mp1)\cosh(b\beta)x+N\mp2=0,\quad\Lambda\rightarrow-0,
\end{equation}
where, for brevity, the following factors have been introduced:
\begin{subequations}
\begin{eqnarray}
x&=&e^{-\mu\beta}e^{-\beta\left/\left(\pi^2|\Lambda|^2\right)\right.}\\
b&=&\frac{4}{\pi^2|\Lambda|^2}e^{-|\Lambda|^{-1}}.
\end{eqnarray}
\end{subequations}
Mathematical solutions of this quadratic with respect to $x$ equation lead to the different physical consequences for the two types of the grand canonical averaging.
\subsubsection{Fermions}\label{Sec_SymmetricFermions1}
\begin{figure}
\centering
\includegraphics[width=\columnwidth]{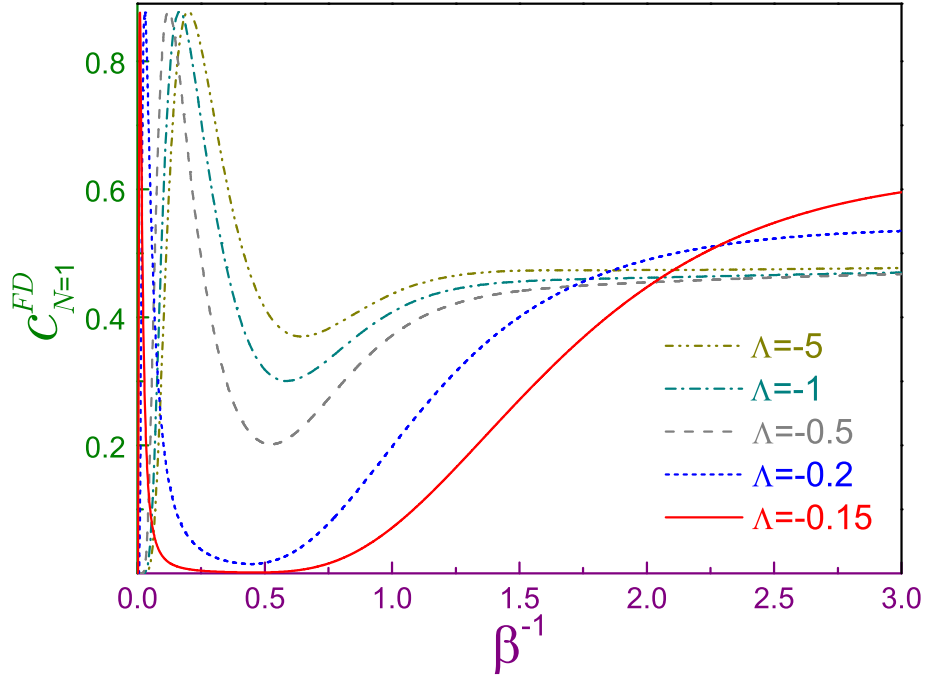}
\caption{\label{SymmetricFDFig1}FD heat capacity $c_1$ of the symmetric QW as function of normalized temperature $\beta^{-1}$.
}
\end{figure}
Physically acceptable solution of Eq.~\eqref{NumberN_2} exists for {\em one} fermion only and it states that the Fermi energy in this regime lies strictly in the middle between the two split-off states:
\begin{equation}\label{FermiSymmetric1}
\mu=-\frac{1}{\pi^2|\Lambda|^2},\quad\Lambda\rightarrow-0.
\end{equation}
Then, the mean energy reads:
\begin{subequations}\label{SymmetricMeanEnCap1}
\begin{align}\label{SymmetricMeanEnCap1_En1}
\left\langle E\right\rangle=-\frac{1}{\pi^2|\Lambda|^2}-b\tanh\frac{b\beta}{2},\quad\Lambda\rightarrow-0,
\intertext{what leads to the following expression for the heat capacity:}
\label{SymmetricMeanEnCap1_Cap1}
c_{N=1}^{FD}=\frac{1}{2}\left(b\beta\sech\frac{b\beta}{2}\right)^2,\quad\Lambda\rightarrow-0.
\end{align}
\end{subequations}
Specific heat reaches its $\Lambda$-independent maximum 
\begin{equation}
c_{max}=\frac{2s}{\sinh s}=0.8785
\end{equation}
at the temperature which very promptly tends to zero with the decreasing Robin length:
\begin{equation}
\beta_{max}=\frac{\pi^2}{4}s|\Lambda|^2e^{-|\Lambda|^{-1}},
\end{equation}
where now the coefficient $s=2.3994$ is a solution of equation:
\begin{equation}
x\sinh x-2\cosh x-2=0.
\end{equation}
The behavior near the maximum is described by
\begin{subequations}\label{SymmetricFDAsymptote1}
\begin{align}
c_1^{FD}(\beta)&=c_{max}-\frac{8s_1}{\pi^4|\Lambda|^4}e^{-2|\Lambda|^{-1}}\left(\beta-\beta_{max}\right)^2,\nonumber\\
&\left|\beta-\beta_{max}\right|\ll\beta_{max},
\intertext{or, equivalently,}
c_1^{FD}\left(\beta^{-1}\right)&=c_{max}-\frac{\pi^4}{32}s^4s_1|\Lambda|^4e^{2|\Lambda|^{-1}}\left(\beta^{-1}-\beta_{max}^{-1}\right)^2,\nonumber\\
&\left|\beta^{-1}-\beta_{max}^{-1}\right|\ll\beta_{max}^{-1},
\end{align}
\end{subequations}
where
\begin{equation}
s_1=s\frac{3\sinh s-s(\cosh s-2)}{(1+\cosh s)^2}=0.4392.
\end{equation}
Eqs.~\eqref{SymmetricFDAsymptote1} show a rapid shrink of the width of the resonance at the vanishingly small negative extrapolation lengths, which is vividly exemplified  in Fig.~\ref{SymmetricFDFig1} showing $c_{N=1}^{FD}$ as a function of $T$ at several $\Lambda$. Physical explanation of this phenomenon is the same as for the canonical distribution, SubSec.~\ref{SubSec_SymmetricCanonical}. Also, it is clear why this maximum at the small temperatures exists for $N=1$ only; indeed, for any larger number of fermions the second lowest orbital will be occupied and, accordingly, transitions to it from the ground state will be forbidden.
\begin{figure*}
\centering
\includegraphics[width=1.2\columnwidth]{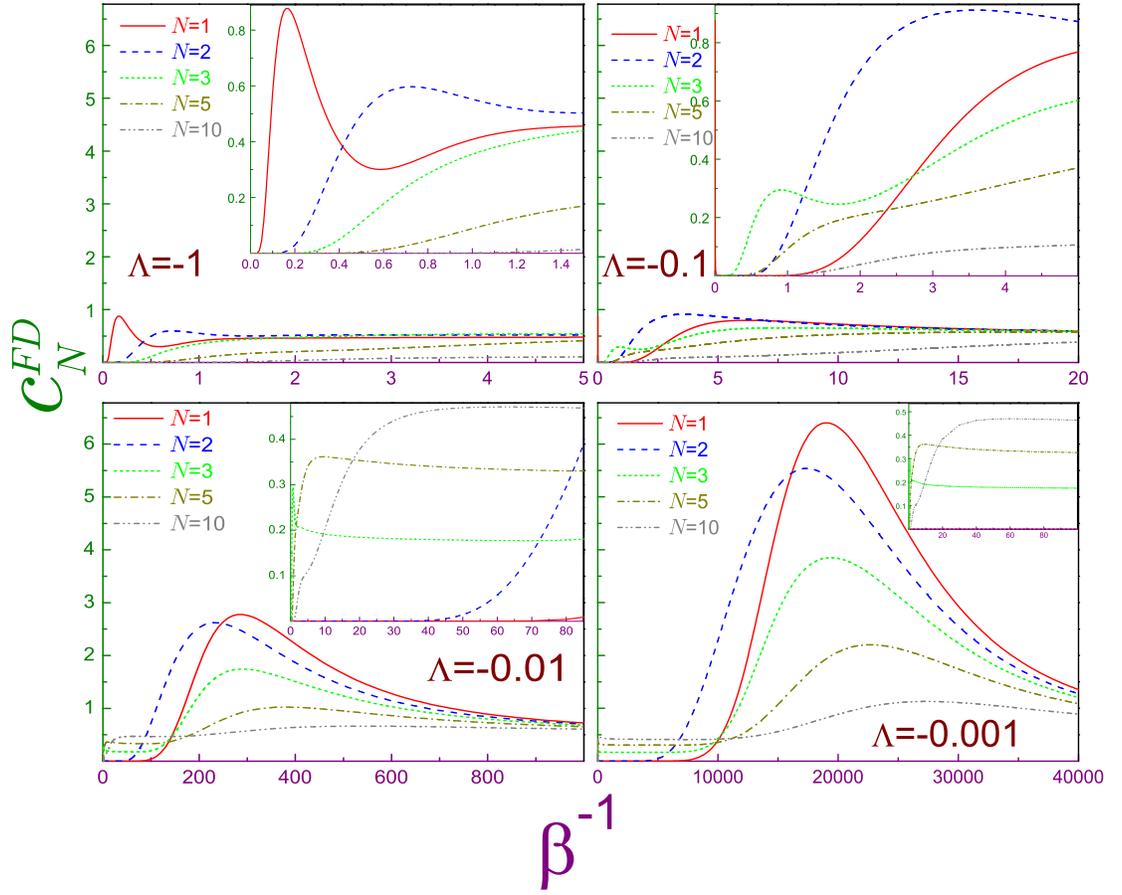}
\caption{\label{SymmetricFDFig2}Fermionic heat capacity per particle $c_N^{FD}$ of the symmetric QW as a function of temperature $\beta^{-1}$ for several extrapolation lengths and number of fermions. The same conventions as in Fig.~\ref{AsymmetricFDFig1} are adopted. The insets each of which has its own vertical and horizontal scales show enlarged views at the small temperatures where the formation of plateaus from Eq.~\eqref{Plateau2} is demonstrated.
}
\end{figure*}

Turning to the discussion of the high-temperature resonances achieved at the small negative extrapolation length, one notices that, similar to the canonical ensemble, they can be mathematically explained with the help of formulas modified from their asymmetric counterparts; for example, first right-hand side items of Eqs.~\eqref{ChemicalPotential2}, \eqref{GrandCanonicalMeanEnergy2}, \eqref{ChemicalPotential3} -- \eqref{GrandCanonicalMeanEnergy4} at $\Lambda\rightarrow-0$ have to be multiplied by two what, for the case of one fermion, leads to a decrease of the maximum and its shift to the warmer temperatures as compared to the asymmetric configuration what was also the case for the canonical distribution. Situation changes for the larger number of particles in the well; say, for $N=2$ both corpuscles reside on the split-off orbitals whereas for $\Lambda_-=-\Lambda_+$ the second electron occupies lowest positive-energy state and, accordingly, its contribution to the total specific heat is much smaller than for its symmetric fellow. This is seen in Fig.~\ref{SymmetricFDFig2} and its comparison with Fig.~\ref{AsymmetricFDFig1}, especially for small $|\Lambda|$; namely, lower panels of Fig.~\ref{SymmetricFDFig2} demonstrate that, contrary to the asymmetric geometry, the heat capacities for one and two fermions do not appreciably differ. Two other features discussed in SubSec.~\ref{Sec_AsymmetricFermions1} are also altered by the existence of the {\em two} split-off levels; namely, at $\Lambda\rightarrow-0$ the maximum corresponding to the interaction at $N=1$ of the two lowest Dirichlet states \cite{Olendski22}, which takes place at the moderate temperature $\beta_{max}^{-1}\lesssim1$, is observed for $N=3$ and its magnitude is one third of that for the hard-wall QW with one fermion. Formation of this resonance is depicted in the insets to panels in Fig.~\ref{SymmetricFDFig2}. In addition, for $N\leq2$ the plateau of the heat capacity is very close to zero whereas for the larger number of fermions in the system it approaches at $\Lambda\rightarrow-0$ the values
\begin{equation}\label{Plateau2}
c_{N_{pl}}=\frac{1}{2}\frac{N-2}{N},\quad N=2,3,\ldots
\end{equation}
[cf. with Eq.~\eqref{Plateau1} for the asymmetric QW].
\subsubsection{Bosons}\label{Sec_SymmetricBosons1}
\begin{figure*}
\centering
\includegraphics[width=1.2\columnwidth]{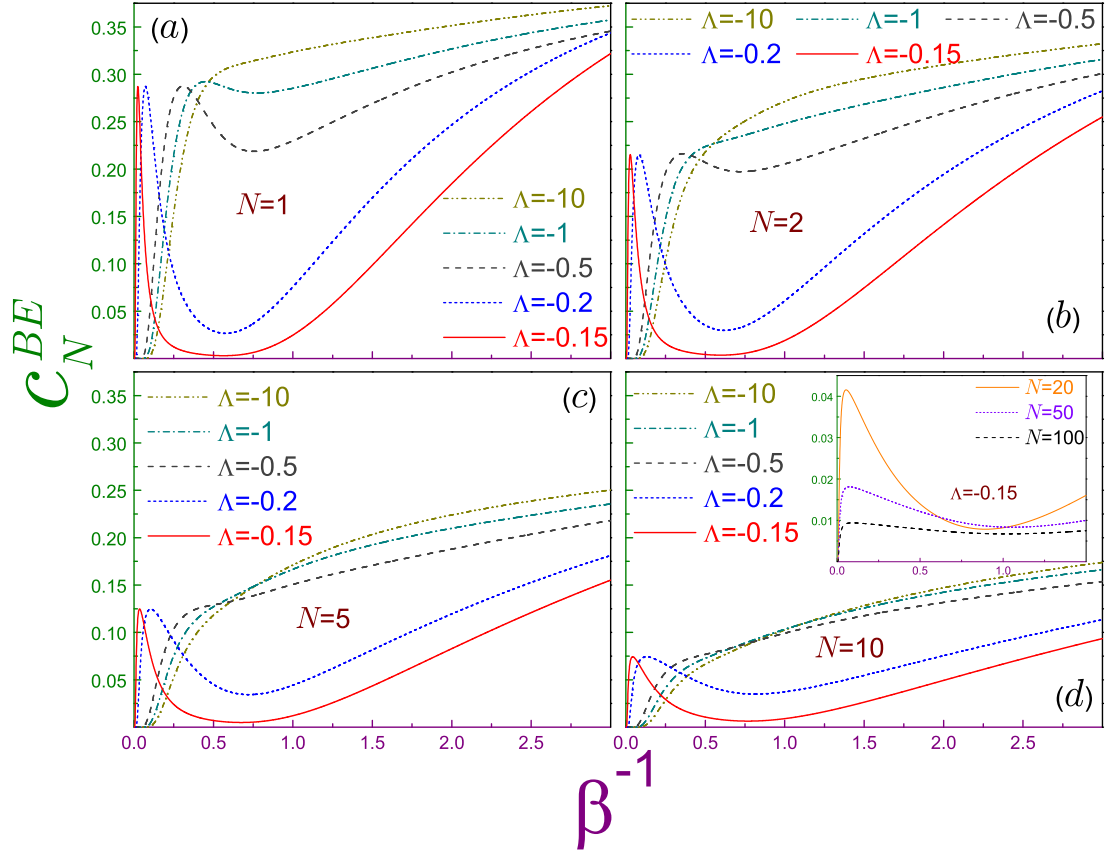}
\caption{\label{SymmetricBEFig1}Bosonic heat capacity per particle $c_N^{BE}$ as a function of small and moderate temperature $\beta^{-1}$ for several negative extrapolation lengths where panel (a) is for one particle, (b) for two corpuscles, (c) for five and (d) for ten bosons. Dash-dot-dotted lines correspond to $\Lambda=-10$, dash-dotted curves are for $\Lambda=-1$, dashed ones -- for $\Lambda=-0.5$, dotted lines -- for $\Lambda=-0.2$, and solid curves are for $\Lambda=-0.15$. Inset in panel (d) shows specific heat dependencies at $\Lambda=-0.15$ and several number of particles where solid line is for $N=20$, dotted curve -- for $N=50$ and dashed one -- for $N=100$. Note different vertical and horizontal scales in the main panels and the inset.
}
\end{figure*}

\begin{figure*}
\centering
\includegraphics[width=1.2\columnwidth]{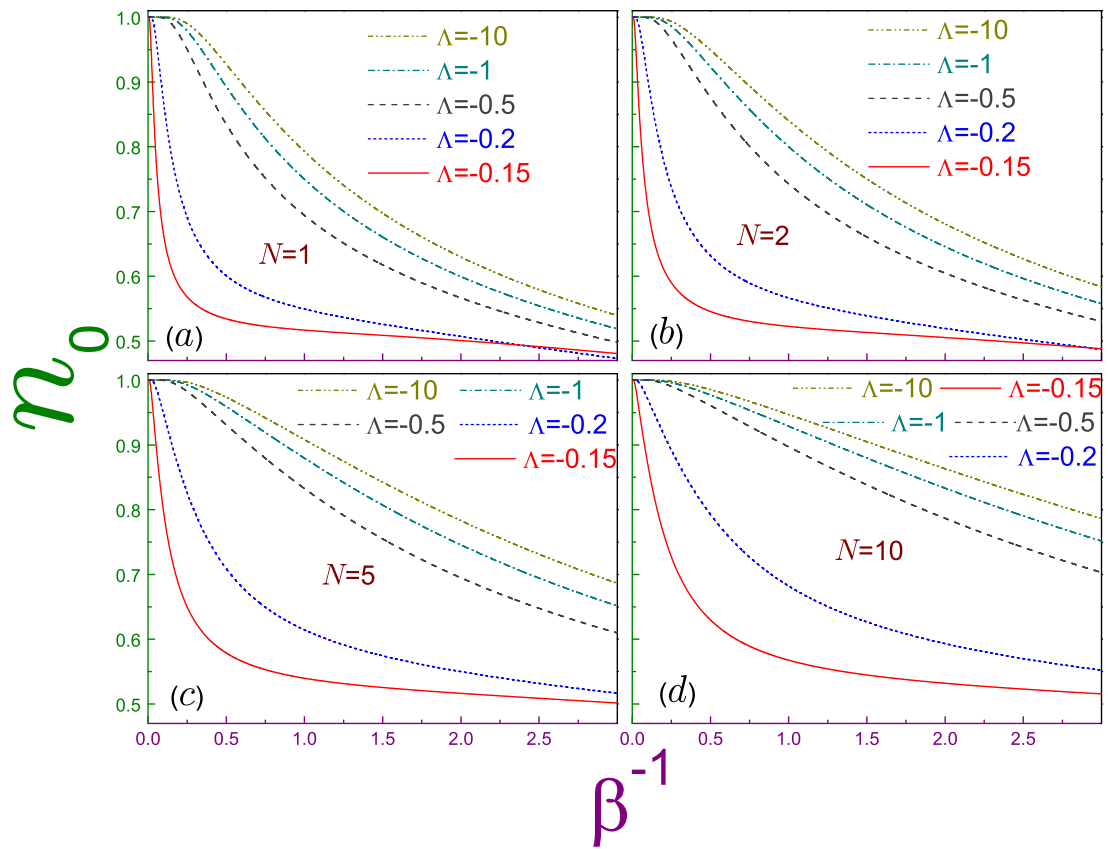}
\caption{\label{SymmetricBEFig2}The same as in Fig.~\ref{SymmetricBEFig1} but for the ground-state population $n_0$.
}
\end{figure*}

Two-level bosonic system represented by Eq.~\eqref{NumberN_2} does not have as simple analytic solution as its $N=1$ fermionic counterpart. Exact results presented in Figs.~\ref{SymmetricBEFig1} and \ref{SymmetricBEFig2} show that now the low-temperature extremum of the heat capacity is formed at the vanishingly small negative Robin parameter for any number of the particles. Similar to the FD statistics, it is characterized by the $\Lambda$-independent magnitude with the corresponding location moving to zero temperature at $\Lambda\rightarrow-0$ whereas the associated half width in the same limit very promptly shrinks. Note that for $N=1$ the BE maximum is more than three times smaller than the FD extremum. Collective behavior of the bosons is exemplified by the fact that adding more corpuscles to the well leads to the decrease of the magnitude of the peak and a formation of the resonant shape of the heat capacity at the shorter absolute values of the Robin parameter; for example, as is seen in the inset of panel (d) of Fig.~\ref{SymmetricBEFig1}, at $N=100$ the $\Lambda=-0.15$ maximum almost completely disappeared. As Fig.~\ref{SymmetricBEFig2} demonstrates, ground-state population $n_0$ for the lesser $|\Lambda|$ descends faster from its zero-temperature unit value what is naturally explained by the smaller gap between the ground and first-excited orbitals. In the extreme limit of $\Lambda\rightarrow-0$ one has two-level system with the very tiny energetic difference between them; as a result, the ground-state population will tend to $1/2$. As Fig.~\ref{SymmetricBEFig2} depicts, a statistical correlation of the motion between several (or many) bosons results in a less precipitous $n_0-T$ dependence at the larger $N$.

\begin{figure}
\centering
\includegraphics[width=1\columnwidth]{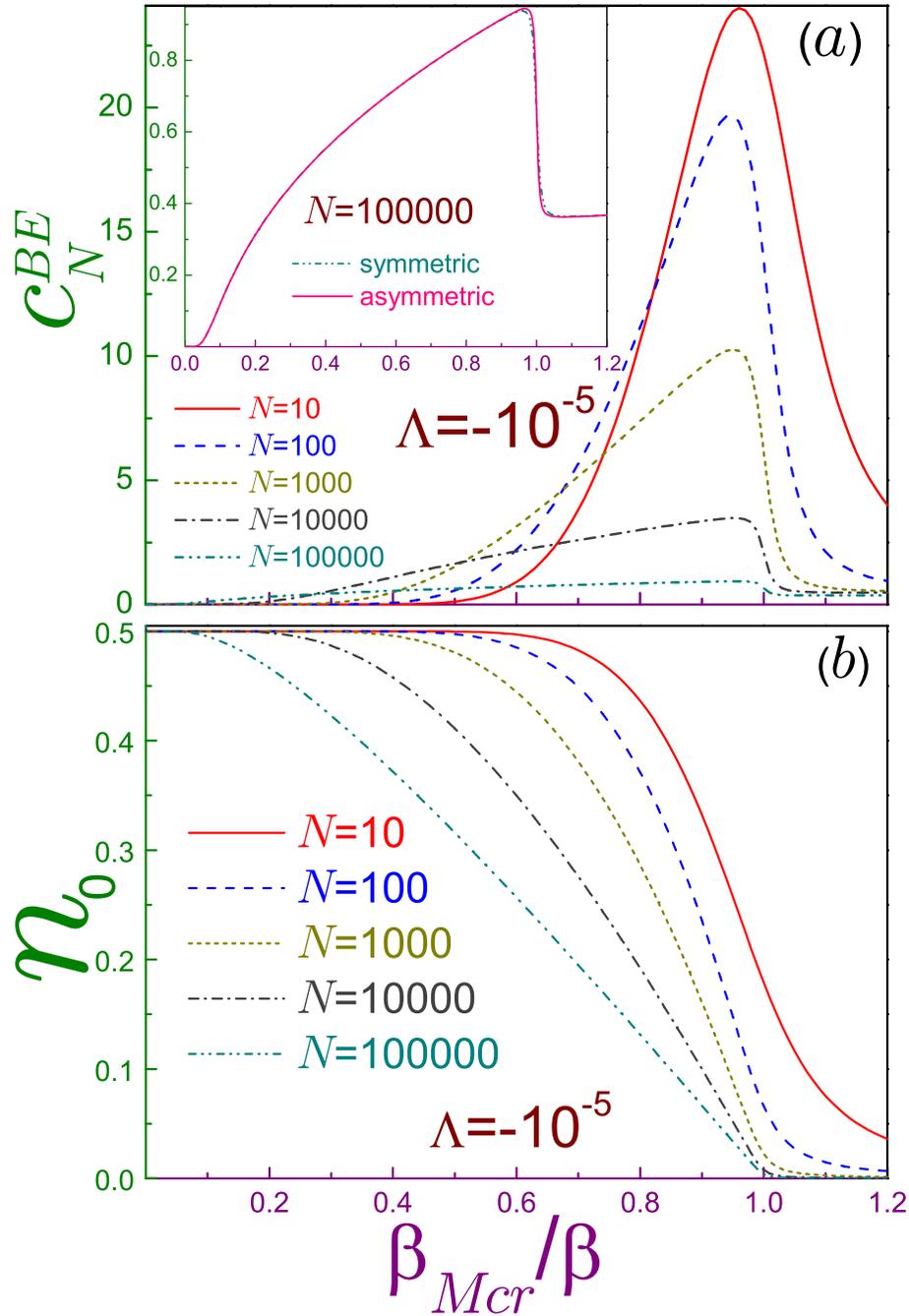}
\caption{\label{SymmetricBEFig3}Bosonic (a) specific heat per particle $c_N$ and (b) ground-state occupation $n_0$ as functions of the temperature [in terms of the modified critical temperature $\beta_{Mcr}^{-1}$, Eq.~\eqref{CriticalTemp5}] at $\Lambda=-10^{-5}$ where solid lines are for $N=10$, dashed curves -- for $N=100$, dotted -- for $N=1000$, dash-dotted line are for $N=10000$, and dash-dot-dotted curves describe dependencies for $N=100000$. Inset in panel (a) compares heat capacities of the symmetric and asymmetric QWs for $N=100000$ where asymmetric function for $\Lambda=10^{-5}$ is borrowed from Fig.~\ref{AsymmetricBEFig3}.
}
\end{figure}

For correct description of the high-temperature resonances of the heat capacity of the symmetric QW, one needs to amend the definition of the critical temperature introduced in SubSec.~\ref{SubSec_AsymmetricBE}; namely, for the extremely small negative Robin parameter, Eq.~\eqref{CriticalTemp1} yields:
\begin{eqnarray}
&&\left[\exp\!\left(\frac{8}{\pi^2|\Lambda|^2}e^{-|\Lambda|^{-1}}\beta_{cr}\right)-1\right]^{-1}\nonumber\\
\label{CriticalTemp3}
&+&\sum_{n=1}^\infty\frac{1}{\exp\!\left[\left(n^2+\frac{1}{\pi^2|\Lambda|^2}\right)\beta_{cr}\right]-1}=N.
\end{eqnarray}
Under the assumption of the small critical temperature, which is confirmed below, Eq.~\eqref{CriticalTemp4}, one discards the series in Eq.~\eqref{CriticalTemp3} and gets:
\begin{equation}\label{CriticalTemp4}
\beta_{cr}=\frac{\pi^2|\Lambda|^2}{8}e^{|\Lambda|^{-1}}\ln\frac{N+1}{N},\quad\Lambda\rightarrow-0.
\end{equation}
The exponential decrease of $T_{cr}$ in this regime is due to the fact that the two lowest split-off levels are almost degenerate with the separation between them described by Eq.~\eqref{SymmetricCanonicalAsymptote4}. Because of this degeneracy, it is reasonable to introduce a modified critical temperature $\beta_{Mcr}^{-1}$ that describes the situation when the chemical potential is locked on the  higher-lying negative-energy orbital, $\mu=E_1$, and no bosons occupying the two split-off levels, $N_0=N_1=0$:
\begin{equation}\label{CriticalTemp5}
\sum_{n=2}^\infty\frac{1}{e^{(E_n-E_1)\beta_{Mcr}}-1}=N,\quad\Lambda\rightarrow-0.
\end{equation}
Then, it possesses the same properties as its counterpart for the asymmetric QW; in particular, Eqs.~\eqref{CriticalTemp2} are valid for it too (of course, with the change of the limiting point to the negative zero). Fig.~\ref{SymmetricBEFig3} shows heat capacity $c_N$ and ground-state population $n_0$ dependencies on the temperature for the small negative extrapolation length $\Lambda=-10^{-5}$. A formation of the asymmetric cusp-like structure of the $c_N-T$ characteristics at $\beta=\beta_{Mcr}$ is clearly seen. A comparison between symmetric and asymmetric structures presented in the inset of panel (a) demonstrates that the choice of the modified critical temperature from Eq.~\eqref{CriticalTemp5} is a correct one: both heat capacities are practically the same with the tiny deviations in the region close to modified $\beta_{Mcr}^{-1}$ or regular $\beta_{cr}^{-1}$ critical temperatures. As discussed above, ground-state population for the two-level BE system at the temperatures greater than the energy difference between the orbitals is equal to $1/2$. Since for the smaller negative extrapolation lengths this asymptote is achieved at the colder temperatures, the very precipitous drop of $n_0$ from unity at $T=0$ to one half is not resolved for the horizontal scale of the figure. Depletion of the two split-off levels at the growth of the temperature depends on the number of bosons in the QW and at quite large $N$ the ground-state population at the temperatures greater than the modified critical one is very small what is another indication of the transition from the BE phase to the normal state.

\section{Concluding remarks}\label{Conclusions}
If the lowest energy of the purely discrete, countably infinite spectrum of the quantum particle in the confining potential is split-off from its higher-lying neighbors in such a way that the ratio $\Delta_{n+1}/\Delta_n$ [with $\Delta_n$ being a difference between energies of the orbital with the principal index $n$ and the ground state, Eq.~\eqref{Spacing2}] stays approximately the same for many excited levels, the heat capacity of this system will exhibit as a function of temperature a pronounced maximum whose magnitude will increase at the above-mentioned ratio decreasing. This physical phenomenon, which was predicted first, to the best of the author' knowledge, in Ref.~\cite{Olendski33}, was theoretically reconfirmed here for the Robin QW with either one or both of its walls represented by the negative extrapolation length whose absolute value describes the attractiveness of the corresponding interface. Qualitative explanation of this giant enhancement of the specific heat is quite simple: the quantum particle, which at the low temperature resides on the ground orbital only, can be promoted by the growing thermal quantum $k_BT$ (we switch back to the normal, unnormalzied units) with about the same probability not only to the first excited state but to the huge number of other orbitals yielding a drastic increase of the specific heat. Similar to the attractive quantum wall in the electric field \cite{Olendski33}, the property persists for any type of the statistical averaging but at the number of the corpuscles $N$ in the system increasing it undergoes different quantitative and qualitative changes; in particular, the magnitude of the maximum decreases if the additional fermions are added to the system and, depending on $\Delta_1/N$, practically disappears whereas for the particles obeying the BE statistics the resonance transforms into the highly asymmetric cusp-like dependence that is a manifestation of the transition from the condensate phase into the normal state. If the two lowest energies are separated from the rest of the spectrum by the large energy gap, at the low temperatures they can be considered as a two-level system and transitions between them yield extra maximum whose location is shifted closer to $T=0$ with $\Delta_1$ decreasing and whose magnitude tends to the constant value independent of this difference.

Analysis above was restricted to the 1D geometry. It is easy to extend it to higher dimensions; namely, for the canonical ensemble, the partition functions $Z^{lD}$ and mean energies $\left\langle E^{lD}\right\rangle$ of the $l$-dimensional QW, $l=1,2,\ldots$, with the same width and the same BC distribution in each direction are, respectively:
\begin{subequations}\label{Ldimensional1}
\begin{eqnarray}\label{Ldimensional1_Z1}
Z^{lD}&=&\left(Z^{1D}\right)^l\\
\label{Ldimensional1_E1}
\left\langle E^{lD}\right\rangle&=&l\left\langle E^{1D}\right\rangle,
\end{eqnarray}
\end{subequations}
where $Z^{1D}$ and $\left\langle E^{1D}\right\rangle$ are the 1D quantities discussed above. Eq.~\eqref{Ldimensional1_E1} shows, in particular, that the canonical heat capacity for the $l$ dimensions will reach its $l$th times higher maximum at the same temperature as its 1D counterpart. It is also clear that, for example, for the cubic Robin quantum dot, $l=3$, the specific heat at high temperatures will approach the value of $3/2$, as expected. Grand canonical ensemble requires more careful analysis; for example, for $l=2$ and antisymmetric geometry Eq.~\eqref{NumberN_1} will transform to
\begin{eqnarray}
N&=&\frac{1}{e^{-\mu\beta}e^{-2\beta\left/\left(\pi^2\Lambda^2\right)\right.}\pm1}+2\sum_{n=1}^\infty\frac{1}{e^{-\left[\mu+1\left./\left(\pi^2\Lambda^2\right)\right.\right]\beta}e^{\beta n^2}\pm1}\nonumber\\
\label{LdimensionalNumberN_1}
&+&\sum_{n_x=1}^\infty\sum_{n_y=1}^\infty\frac{1}{e^{-\mu\beta}e^{\beta\left(n_x^2+n_y^2\right)}\pm1}.
\end{eqnarray}
Compared to its 1D counterpart, Eq.~\eqref{ChemicalPotential2}, this expression contains more terms including double series. However, in our range of interest, using the assumptions and approximations employed in Sec.~\ref{SubSec_AsymmetricGrandCanonical}, Eq.~\eqref{LdimensionalNumberN_1} and accompanying formula for the mean energy apparently can be simplified yielding analytic results similar to those developed above. For the cubic quantum dot the number of terms in the corresponding equations increases even more and they will contain triple infinite series but the system looks again analytically and numerically tractable.

For practical applications, the most important is the fact that the temperature of the $T$-dependent maximum gets hotter when the extrapolation length (for the asymmetric QW) or the absolute value of the negative Robin distance (for symmetric structure) decreases. It allows, for example, to keep the bosonic condensate phase at the higher $T$. For the single surface, it was conjectured that the model of the attractive interface describes piecewise continuous potentials \cite{Pazma1,Fulop1,Olendski33} that can be grown by the modern semiconductor technologies where the charge carriers (electrons or holes) are described by the FD statistics. The same procedure can be appropriately modified to create the QW with two surfaces characterized by the negative extrapolation length. Designing the corresponding bosonic structure with the variable Robin distance will allow to experimentally check the evolution of the heat capacity predicted above.

\section{Acknowledgments}
Research was supported by SEED Project No. 1702143045-P from the Research Funding Department, Vice Chancellor for Research and Graduate Studies, University of Sharjah.

\bibliographystyle{model1a-num-names}
 
\end{document}